\documentclass{SciPost}

\hypersetup{
    colorlinks,
    linkcolor={red!50!black},
    citecolor={blue!50!black},
    urlcolor={blue!80!black}
}

\DeclareSymbolFont{usualmathcal}{OMS}{cmsy}{m}{n}
\DeclareSymbolFontAlphabet{\mathcal}{usualmathcal}

\fancypagestyle{SPstyle}{
\fancyhf{}
\lhead{\colorbox{scipostblue}{\bf \color{white} ~SciPost Physics }}
\rhead{{\bf \color{scipostdeepblue} ~Submission }}

\fancyfoot[C]{\textbf{\thepage}}
}

\usepackage[utf8]{inputenc}
\usepackage{amsmath,amssymb,bbm,dutchcal}
\usepackage{amsthm}
\usepackage{graphicx}
\usepackage{xcolor}
\usepackage{mathtools}
\usepackage{tikz}
\usepackage{float}
\usepackage{enumerate}

\newcommand{\ii}{{\rm i}}
\newcommand{\1}{\mathbbm{1}}
\newcommand{\tr}{\mathrm{tr} \,}
\newcommand{\ev}[1]{\left\langle #1 \right\rangle}
\newcommand{\sprod}[2]{\left\langle #1, #2 \right\rangle}
\newcommand{\sprode}[2]{\left\langle\!\left\langle #1, #2 \right\rangle\!\right\rangle}
\newcommand{\abs}[1]{\left\lvert#1\right\rvert}
\newcommand{\norm}[1]{\left\lVert#1\right\rVert}

\newcommand{\mev}[2]{\langle #1 | #2 | #1 \rangle}
\newcommand{\dd}[1]{\, \mathrm{d}#1}
\newcommand{\e}[1]{{\rm e}^{#1}}
\newcommand{\coloneq}{\coloneqq}

\newcommand{\mzz}[1]{{#1}}
\newcommand{\diff}[1]{#1}
\newcommand{\diffa}[1]{{#1}}

\begin{document}

\title{Ruelle-Pollicott resonances of diffusive U(1)-invariant qubit circuits}
\author{Urban Duh}
\author{Marko Žnidarič}
\date{\today}

\pagestyle{SPstyle}

\begin{center}{\Large \textbf{\color{scipostdeepblue}{
Ruelle-Pollicott resonances of diffusive U(1)-invariant qubit circuits\\
}}}\end{center}

\begin{center}\textbf{
Urban Duh\textsuperscript{$\star$} and
Marko Žnidarič
}\end{center}

\begin{center}
Physics Department, Faculty of Mathematics and Physics, University of Ljubljana, 1000 Ljubljana, Slovenia
\\[\baselineskip]
$\star$ \href{mailto:urban.duh@fmf.uni-lj.si}{urban.duh@fmf.uni-lj.si}
\end{center}

\section*{\color{scipostdeepblue}{Abstract}}
\textbf{\boldmath{%
  We study Ruelle-Pollicott resonances of translationally invariant
  magnetiza\-tion-conserving qubit circuits via the spectrum of the
  quasi-momentum-re\-solved truncated propagator of extensive observables.
  Diffusive transport of the conserved magnetization is reflected in the
  Gaussian quasi-momentum $k$ dependence of the leading eigenvalue
  (Ruelle-Pollicott resonance) of the truncated propagator for small $k$. This,
  in particular, allows us to extract the diffusion constant. For large $k$, the
  leading Ruelle-Pollicott resonance is not related to transport and governs the
  exponential decay of correlation functions. Additionally, we conjecture the
  existence of a continuum of eigenvalues below the leading diffusive resonance,
  which governs non-exponential decay, for instance, power-law hydrodynamic
  tails. We expect our conclusions to hold for generic systems with exactly one
  U(1) conserved quantity.
}}

\vspace{\baselineskip}

\noindent\textcolor{white!90!black}{%
\fbox{\parbox{0.975\linewidth}{%
\textcolor{white!40!black}{\begin{tabular}{lr}%
  \begin{minipage}{0.6\textwidth}%
    {\small Copyright attribution to authors. \newline
    This work is a submission to SciPost Physics. \newline
    License information to appear upon publication. \newline
    Publication information to appear upon publication.}
  \end{minipage} & \begin{minipage}{0.4\textwidth}
    {\small Received Date \newline Accepted Date \newline Published Date}%
  \end{minipage}
\end{tabular}}
}}
}


\pagebreak
\tableofcontents
\pagebreak

\section{Introduction}

One of the reasons why physics is so successful in describing nature is because
oftentimes simple descriptions suffice. This is so since even simple rules,
given for instance by a Hamiltonian, can lead to complicated behavior. How and
which simple rules lead to complex behavior is studied within the theory of
dynamical systems. Some systems (rules) lead to predictable regular dynamics,
the extreme case being integrable systems, while others lead to complex behavior,
the extreme case being chaotic systems for which predicting the future for a
specific initial condition becomes impossible due to exponential complexity
growth~\cite{ottChaosDynamicalSystems2002}.

In many-body quantum systems, complexity can grow exponentially also with the
system size. Such exponentially growing Hilbert space size is what allows for
outperformance of quantum devices over classical, and indeed, one of the premier
uses of present day digital quantum computers and analog simulators is to
simulate quantum many-body systems~\cite{Daley}. An interesting question is:
What new effects can arise from the combination of exponential complexity due to
chaos and that from the many-body nature? To begin with, it is not entirely
clear how to define or quantify quantum chaos in many-body quantum systems. One
can try to use theoretically appealing concepts, like dynamical entropies, which
are difficult to use in practice, or one can use a concept that is easy to study
numerically, for instance one of the various spectral quantifiers~\cite{Haake}.
Both have a drawback that in a many-body system they are hardly observable
quantities.

One can, however, abandon theoretical elegance for practical relevance and try to
directly study measurable objects. \diffa{An example is the correlation function}
\begin{equation}
    C_{AB}(t) \coloneq \ev{A(t) B(0)} - \ev{A(t)} \ev{B(0)}, \label{eq:cf}
\end{equation}
where $\ev{\bullet}$ denotes the average over an appropriate ensemble. A
defining property of mixing systems is that correlation functions decay to zero,
while for chaotic systems we usually expect more, namely, that $C_{AB}(t)$ decay
exponentially at long enough times, $C_{AB}(t) \sim \e{-\nu t}$. Time-dependent
correlation functions specifically allow us to study dynamics, and through
linear response also get access to non-equilibrium properties. A prime example
is the transport of globally conserved quantities, like energy, charge, or
magnetization.

While many different methods have been proposed and employed to study transport
in many-body quantum systems, for a review see
Ref.~\cite{bertiniFinitetemperatureTransportOnedimensional2021}, here we propose
and test a new one that is based on the dynamics of operators. It directly works
in the thermodynamic limit, i.e., in infinite-sized systems, with the only
approximation being truncating the dynamics to operators with a finite local
support size. It builds on a decades-old idea of Ruelle-Pollicott (RP)
resonances~\cite{ruelleResonancesChaoticDynamical1986,pollicottRateMixingAxiom1985,gaspardChaosScatteringStatistical1998}
that we upgrade and adapt to translationally invariant many-body quantum
circuits, so that it allows for easy an calculation of the transport dynamical
exponent and phenomenological constants, for instance, the diffusion constant.

\subsection{Ruelle-Pollicott resonances}
\label{sec:rp_res}

The idea of RP resonances is to get the decay rate $\nu$ of the asymptotic
exponential decay of correlation functions,
\begin{equation}
    C_{AB}(t) \sim \e{-\nu t} = \lambda^t, \label{eq:rp_res}
\end{equation}
where we introduced $\lambda \coloneq \e{-\nu}$. While specific observables can
decay with their own rate, there is the slowest decay rate $\nu$, whose inverse
is the system's relaxation time. It was shown in
Refs.~\cite{pollicottRateMixingAxiom1985,ruelleResonancesChaoticDynamical1986}
that in a class of strongly chaotic classical systems $\lambda$ does not depend
on the chosen observable, but is rather an intrinsic property of the system. In
such cases, $\lambda$ is referred to as the (leading) RP resonance of the
system. While one could get $\lambda$ by simply calculating the appropriate
correlation function, the idea is to get $\lambda$ directly from the generator of
dynamics (after all, $\lambda$ is an inherent property of the system, not of the
observables).

With classical systems in mind, the starting point is the propagator $\mathcal
U$, either the Frobenius-Perron propagator of densities, or the Koopman propagator
of observables~\cite{gaspardChaosScatteringStatistical1998,Braun}. $\mathcal U$
is unitary in the appropriate space (e.g., in the space of $L^2$ functions), its
eigenvalues lying on the unit circle. One can, however, find poles in the
analytic continuation of the resolvent $\mathcal R(z) \coloneq 1/(z - \mathcal U)$
inside the unit circle. It was shown in specific
systems~\cite{hasegawaUnitarityIrreversibilityChaotic1992} that they correspond
exactly to RP resonances, the largest being the mentioned $\lambda$.

Performing the analytic continuation for generic systems is not feasible. A more
practical approach, which can be carried out either analytically or numerically,
is introducing some kind of dissipation or coarse-graining (i.e., a
non-unitarity) in our system. This leads to a non-unitary propagator $\mathcal
U(\varepsilon)$, where $\varepsilon$ measures the strength of the non-unitarity.
The eigenvalues of $\mathcal U(\varepsilon)$ now lie inside the unit circle and,
analogous to the RP resonances in the unitary case, govern the decay of generic
correlation functions. It is conjectured that in the unitary limit $\varepsilon
\to 0$ the naive expectation that the eigenvalues uniformly converge
to the unit circle is not correct -- they converge to a smaller radius
inside the unit circle, which corresponds to an RP resonance. This procedure was
used in classical systems, with the non-unitarity being either white
noise~\cite{gaspardSpectralSignaturePitchfork1995} or a phase space
coarse-graining procedure~\cite{weberFrobeniusperronResonancesMaps2000}.

Less is known about RP resonances of quantum systems. Initially, they have been
studied in the semi-classical
regime~\cite{panceQuantumFingerprintsClassical2000,
manderfeldClassicalResonancesQuantum2003, garcia-mataChaosSignaturesShort2018}.
More recently, weak Lindblad dissipation (an analog of noise in classical
systems) has been suggested~\cite{moriLiouvilliangapAnalysisOpen2024} as a
method of extracting $\lambda$ via the Liouvillian gap in the limit of infinite
system size and zero noise, with the results for a few specific solvable
models~\cite{jacobySpectralGapsLocal2024, zhangThermalizationRatesQuantum2024,
yoshimuraTheoryIrreversibilityQuantum2025} agreeing with $\lambda$ calculated
directly.

We shall use a different path by considering the Heisenberg propagator of
observables $\mathcal U(A) \coloneq U^\dagger A U$, where $U$ is the unitary
propagator, truncating $\mathcal U$ to an appropriate operator subspace.
In this setting, the RP resonance is expected to correspond to the leading eigenvalue
of $\mathcal U$. The
method was first used on translationally invariant local operators in
Ref.~\cite{prosenRuelleResonancesQuantum2002} (which can be thought of as akin to
a phase space coarse-graining) in the kicked Ising
model~\cite{prosenRuelleResonancesQuantum2002,
prosenRuelleResonancesKicked2004,prosenChaosComplexityQuantum2007}, and more
recently followed by additionally resolving the
quasi-momentum~\cite{znidaricMomentumdependentQuantumRuellePollicott2024}. We
shall use this quasi-momentum-resolved truncated propagator. The truncated
propagator can be very handy not just in studies of chaotic systems, but more
broadly, for instance, in identifying unknown constants of motion in cellular
automata~\cite{katjaphd,sharipovErgodicBehaviorsReversible2025}, or discovering
that all homogeneous U(1) conserving quantum circuits are
integrable~\cite{znidaricIntegrabilityGenericHomogeneous2024,kpzwall}.

We note that the recently much discussed Krylov space method can be viewed as a
particular way of truncation (in powers of time in the Taylor expansion, i.e.,
nested commutators) and can thus be related to RP
resonances~\cite{Lychkovsky25,Buca}.

\subsection{Summary of results}

\begin{figure}[t!]
    \centering
    \raisebox{135pt}{(a)}
    \includegraphics[width=0.42\linewidth]{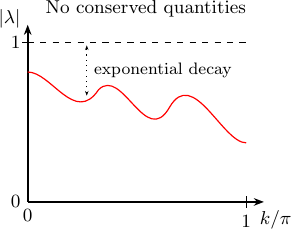}
    \raisebox{135pt}{(b)}
    \includegraphics[width=0.42\linewidth]{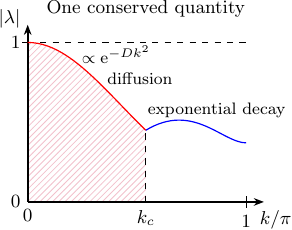}
    \caption{Diagram of the \diffa{$k$-dependent} RP resonance spectrum in
    \diffa{homogeneous} systems without (a), and with one conserved
    quantity (b). Labels in figure (b) show which parts of the spectrum are
    responsible for which physical behavior of the system. \diffa{The shaded region in
    figure (b) denotes the location of conjectured continuums of RP resonances,
    which we cannot observe numerically with the methods used.} We show only
    positive $k$ since $|\lambda(k)| = \abs{\lambda(-k)}$, for details see
    Sec.~\ref{sec:trunc}. }
    \label{fig:rp_diagram}
\end{figure}

In this work we study RP resonances of systems with one conserved quantity. We
focus on many-body quantum circuits with a local U(1) conserved quantity and
local gates. Such systems, with their associated transport, are of
direct experimental interest for NISQ machines~\cite{john,Maruyoshi,google}, and
are the simplest theoretical setting with a conserved
quantity\footnote{Implementation of the truncated propagator method is simpler
for quantum circuits compared to Hamiltonian systems due to a sharp light-cone.
In addition, one has conserved magnetization with a 1-local density instead of
\diffa{conserved} energy with a 2-local density.}. Our generic conclusions are demonstrated on
magnetization-conserving circuits with 3-site interaction and 3-site
translational invariance.

As mentioned, we use the truncated operator
propagator~\cite{prosenRuelleResonancesQuantum2002} in its
quasi-momentum-resolved
form~\cite{znidaricMomentumdependentQuantumRuellePollicott2024}, which allows us
to extract transport properties from the momentum dependence of RP resonances.
Some work has been done on the connection between RP resonances and diffusion in
classical
systems~\cite{gaspardDiffusionEffusionChaotic1992,khodasRelaxationInvariantDensity2000},
and conserved quantities have been briefly mentioned in previous Lindblad weak
dissipation
works~\cite{moriLiouvilliangapAnalysisOpen2024,yoshimuraTheoryIrreversibilityQuantum2025},
as well as for the truncated propagator of the kicked Ising model in a
prethermalized regime (or the Hamiltonian limit) in a recent
preprint~\cite{prethermal}. \mzz{We also note that a somewhat related idea of
dropping operators with many non-identity Pauli operators has been used in
Ref.~\cite{PhysRevB.105.075131}, where the authors introduce a numerical
Lindblad evolution method that aims to reach the long-time hydrodynamic
behavior.}

We provide the first comprehensive study of the $k$-dependence of RP resonances,
showing that (i) the quasi-momentum dependence of the leading eigenvalue
of the truncated propagator and thus the leading RP resonance
$\lambda_1(k)$ is in general non-trivial (Fig.~\ref{fig:rp_diagram}a), even in a
system without any conserved quantities where $|\lambda_1(k)|$ is gapped away
from $1$. For instance, the decay need not be always the slowest at $k=0$, i.e.,
for translationally invariant observables. (ii) In systems with conserved
magnetization, one expectedly has $\lambda_1(0)=1$ with the corresponding
eigenvector being the conserved magnetization, and, more interestingly, from
$\lambda_1(k)$ for small $k$ one can extract the transport dynamical exponent,
and, specifically in the case of diffusion, the value of the diffusion constant
\diffa{from} $|\lambda_1(k)| = \e{-Dk^2}$ (Fig.~\ref{fig:rp_diagram}b). At large $k$,
the diffusive nature of the leading RP resonance ends due to an eigenvalue
crossing at $k_c$, resulting in an exponential decay of generic observables with
sufficiently large quasi-momentum $k$. (iii) We discuss and conjecture an
RP continuum below $|\lambda_1(k)|$
(Fig.~\ref{fig:rp_diagram}b) that is related to non-exponential decay of
correlation functions and corrections to diffusive transport.

\section{Truncated quasi-momentum-dependent propagator of extensive observables}
\label{sec:trunc_prop}

In this section, we describe the quasi-momentum-dependent propagator of local
observables, first introduced for $k = 0$ in
Ref.~\cite{prosenRuelleResonancesQuantum2002} and extended to any $k$ in
Ref.~\cite{znidaricMomentumdependentQuantumRuellePollicott2024}. We are
considering qubit circuits with propagators for $1$ unit of time $U$ that have
translational invariance by $s$ sites, and that can, due to a sharp light-cone,
spread local operators supported on $r$ consecutive sites by at most $\delta r$
sites at each edge. For instance, a brickwall circuit composed of all the same
gates (Fig.~\ref{fig:rp_no_sym}a) is invariant under translation by $s=2$ sites
and has $\delta r=2$, while a 3-layered brickwall circuit with 3-qubit gates
(Fig.~\ref{fig:rp_sym}a) has $s=3$ and $\delta r=6$. Previous
implementations~\cite{prosenRuelleResonancesQuantum2002, prosenRuelleResonancesKicked2004, prosenChaosComplexityQuantum2007, znidaricMomentumdependentQuantumRuellePollicott2024}
were for the kicked Ising model that has $s = 1$ and $\delta r = 1$, or for the
(2-layer) brickwall circuits~\cite{kpzwall,znidaricIntegrabilityGenericHomogeneous2024}. We
generalize it to work for translationally invariant quantum circuits with any
geometry, i.e., any $s$ and $\delta r$.

\diffa{First,} we express the Heisenberg propagator of observables
\begin{equation}
    \mathcal U(A) \coloneq U^\dagger A U
\end{equation}
in \diffa{an operator} basis of an infinite system where the quasi-momentum $k$ is a good quantum
number. We then truncate (i.e., project) the observable $\mathcal U(A)$
back to observables with densities supported on at most $r$ consecutive sites, thus
obtaining a non-unitary truncated propagator $\mathcal U^{(r)}$. In the limit
of the non-unitarity going to zero, that is $r \to \infty$, we expect to extract
RP resonances as the ``frozen'' eigenvalues of the truncated propagator
$\mathcal U^{(r)}$.

\subsection{Brief description}
\label{sec:prop}

In order to study the Heisenberg propagator with good quasi-momentum $k$, we
must first define a basis of operators with good $k$. Any observable in a finite
system of $N$ lattice sites with good quasi-momentum $k$ can be written in the
following way
\begin{equation}
    A = \sum_{j = 0}^{N/s - 1} \e{-\ii kj} \mathcal S^{sj}(a), \label{eq:a_ext_simple}
\end{equation}
where $\mathcal S$ is the 1-site translation super-operator to the right (i.e.,
$\mathcal S(a \otimes \1) = \1 \otimes a$), $a$ is a local operator and the sum
runs over all $j\in \{0, \dots, N/s - 1\}$ possible translations by $s$ sites.
In a finite system, $k$ takes values $\frac{2 \pi s}{N} j$, however, at the end
we will be interested in the thermodynamic limit $N \to
\infty$\footnote{Formally, RP resonances can only exist in systems with
infinite-dimensional Hilbert spaces. In finite systems, correlation functions
always converge to a nonzero constant.}, where we can take $k \in (-\pi, \pi]$. One can check
that $A$ really is an eigenvector of the translation super-operator,
$\mathcal{S}(A) = \e{\ii k} A$. Such $A$ will be referred to as an
\textit{extensive observable}.

We wish to work with local operators $a$ as proxies for extensive observables
$A$. It is thus crucial to restrict ourselves to a space of local operators in
which the correspondence between local operators and extensive observables is
one-to-one. Namely, in Eq.~\eqref{eq:a_ext_simple}, different $a$ can result in
the same $A$. For example, both $a = \sigma^z_1$ and $a =
\frac{1}{2}\left(\sigma^z_1 + \sigma^z_{1 + s}\right)$ represent the same $A$.
Here $\sigma_j^\alpha$ denotes the $\alpha$ Pauli matrix acting on site $j$. To
have a one-to-one correspondence, such translations of components
of $a$ by $s$ sites must be prohibited. This can be done in various ways, we
choose to write the extensive observable in the following form
\begin{equation}
    A = \sum_{m = 0}^{s - 1} \sum_{j = 0}^{N/s - 1} \e{-\ii k j} \mathcal{S}^{sj + m}(a_m), \label{eq:a_ext}
\end{equation}
where $\mathcal S^{sj +m}(a_m)$ can be thought of as operators starting on sites $sj + m$ (i.e., on
all sites $i$, where $i \ (\mathrm{mod}) \ s = m$). In other words, due to
translational invariance by $s$ sites we have a local unit cell of $s$ sites, so
that the sum over $m$ runs over sites within a unit cell, while $j$ runs
over all unit cells. In this formulation it is now easy to see that we have
a one-to-one correspondence between the set of $\{a_m\}_{m = 0}^{s - 1}$ and
the extensive observable $A$ if we enforce that $a_m$ do not contain
components acting trivially on the first site. Explicitly, $a_m$ must be in
the space spanned by the set (basis) $\mathcal P^{(r)}$ of Pauli strings without identities acting on the first site:
\begin{align}
    \mathcal{p}_0 &\coloneq \{\1\} \cup \mathcal{p}, \qquad \text{where} \qquad \mathcal{p} \coloneq \{\sigma^x, \sigma^y, \sigma^z\}, \label{eq:local_b}\\
    \mathcal{P}^{(r)} &\coloneq \left\{p_1 \otimes p_2 \otimes  \cdots \otimes p_r \ | \ p_1 \in \mathcal{p} \ \land \ p_{j \neq 1} \in \mathcal{p}_0\right\}. \nonumber
\end{align}
In this notation, therefore,
\begin{align}
    a_m &\in \mathrm{Span}\left(\mathcal{P}^{(r)}\right) \text{for some $r$}.
\end{align}
Such $a_m$ will be referred to as \textit{local densities}. Additionally, we
define their support to be the smallest $r$ such that $a_m \in
\mathrm{Span}\left(\mathcal{P}^{(r)}\right)$. Analogously, the support of an
extensive observable $A$ is defined to be the smallest $r$, such that $a_m \in
\mathrm{Span}\left(\mathcal{P}^{(r)}\right)$ for all $m \in \{0, \dots, s-1\}$.

The basis of extensive observables can now be written down, one merely takes all
possible local basis elements $\mathcal{P}^{(r)}$ on all possible
starting positions $m$. The basis elements are therefore
\begin{equation}
    B_k^{(m, b)} \coloneq \sum_{j = 0}^{N/s - 1} \e{-\ii k j} \mathcal{S}^{sj + m}(b), \label{eq:ext_b}
\end{equation}
where $b \in \mathcal{P}^{(r)}$ for some $r$. Moreover, $B_k^{(m,b)}$ for all
$k, m$ and $b \in \mathcal{P}^{(r)}$ are an orthonormal basis of the space of
extensive observables w.r.t.\ the \textit{extensive} Hilbert-Schmidt inner
product, $\sprode{A}{B} \coloneq \frac{s}{N 2^N} \tr A^\dagger B$. The number of
basis elements of support $r$ (or less) is $\frac{3s}{4}4^{r}$. Importantly, the
identity operator $\1$ is excluded from this basis. This is convenient, since it
is always a conserved quantity in unitary dynamics and is, therefore, not of interest
in the present application. For details and derivations \diffa{regarding the space
of extensive observables} see Appendix~\ref{app:ext_obs}.

It is now easy to calculate the matrix elements of the Heisenberg propagator
in the basis of extensive observables by simply taking the inner product
\begin{equation}
    \left[\mathcal U_k\right]_{(m, b), (m', b')} = \sprode{B^{(m, b)}_k}{\mathcal U\left(B^{(m', b')}_k \right)}. \label{eq:prop_mel_simple}
\end{equation}
In this context, we call $\mathcal U$ the \textit{propagator of extensive
observables} (i.e., the Heisenberg propagator in a particular basis).

\subsection{Truncation and Ruelle-Pollicott resonances}
\label{sec:trunc}

The Heisenberg propagator in Eq.~\eqref{eq:prop_mel_simple} evaluated on the entire
(infinite-dimensional) Hilbert space is a unitary operator. We wish to
introduce a non-unitarity by restricting ourselves to some subspace. There
are different options, one of the most natural ones is restricting
ourselves to the space of support $r$ defined in Eq.~\eqref{eq:local_b}. Projection of the Heisenberg propagator
to this space results in the \textit{truncated propagator of extensive observables}
$\mathcal U^{(r)}_k$ with matrix elements
\begin{equation}
    \left[\mathcal U_k^{(r)}\right]_{(m, b), (m', b')} = \sum_{j = -1}^1 \e{\ii k j} \sprod{\mathcal S^{sj + m}(b)}{\mathcal U \left(\mathcal S^{m'}(b')\right)}, \label{eq:prop_mel}
\end{equation}
where $b, b' \in \mathcal P^{(r)}$ (\ref{eq:local_b}), $m, m' \in \{0, \dots, s
- 1\}$ and $k \in (-\pi, \pi]$. The RHS is expressed only in terms of local
densities and the angled brackets denote the \textit{local} Hilbert-Schmidt
inner product, $\sprod{a}{b} \coloneq \frac{1}{2^N} \tr a^\dagger b$. Importantly, the truncated propagator is
block-diagonal with blocks indexed by the quasi-momentum $k$. The written
form is valid if $\delta r \leq s$. More generally,
the range of values of $j$ over which one has to sum in Eq.~\eqref{eq:prop_mel}
depends on the spreading $\delta r$ of the circuit. For the derivation and
further details see Appendix~\ref{app:prop}.

The truncated propagator $\mathcal U^{(r)}_k$ is non-unitary and its eigenvalues
lie inside the unit circle. One might think that its eigenvalues smoothly converge
back to the unit circle in the unitary limit $r \to \infty$, but in many cases
this is not true. If the local correlations decay exponentially one expects that
the leading eigenvalue converges (or ``freezes'') to some value inside the unit
circle, which one interprets as the leading RP resonance $\lambda_1(k)$.

For the finite-$r$ truncated propagator \diff{of an infinite circuit
$ N \to \infty$}, assuming it is diagonalizable, we can
\diffa{illustrate} that more explicitly. Let $\lambda_i^{(r)}$ be the eigenvalues of
$\mathcal U_k^{(r)}$ (sorted by decreasing magnitude) and $\mathfrak L_i^{(r)}$/
$\mathfrak R_i^{(r)}$ its left/right eigenvectors. A generic
infinite-temperature correlation function of observables $A$, $B$ with good
quasi-momentum $k$ can then be decomposed as
\begin{align}
    \ev{A(0)B(t)} &= \sprod{A^\dagger}{\mathcal U_k^t(B)} \label{eq:rp_decay_show}\\
    &= \lim_{r \to \infty} \sum_i \left(\lambda_i^{(r)}\right)^{t} \sprod{A^\dagger}{\mathfrak R_i^{(r)}}\sprod{\mathfrak L_i^{(r)}}{B} \nonumber \\
    &\xrightarrow[t \to \infty]{} \lambda_1^{t} \sprod{A^\dagger}{\mathfrak R_1^{(\infty)}}\sprod{\mathfrak L_1^{(\infty)}}{B}. \nonumber
\end{align}
Therefore, \diff{provided that $A, B$ have a finite overlap with} the eigenvectors associated with the \mzz{isolated}
leading \diff{eigenvalue}, their correlation function will decay as
$\abs{\lambda_1}^t$ at long times, which matches with the definition of an
RP resonance in Eq.~\eqref{eq:rp_res}. Here on the LHS $\ev{\bullet} \coloneq \frac{1}{2^N}
\tr(\bullet)$ (i.e., an infinite-temperature average in a finite system) and,
for clarity, $A, B$ are assumed to be traceless. If $A, B$ are not traceless
(i.e., they overlap with the identity, which is also the eigenvalue of $\mathcal
U$), a similar argument can be made for the connected correlation function
defined in Eq.~\eqref{eq:cf}. In all the following sections we shall discuss
connected infinite-temperature autocorrelation functions, we will usually
shorten our language and simply say correlation functions. A similar procedure can
also be done at finite temperature or chemical potential by introducing an
appropriate inner product with a non-trivial measure.

If a circuit has a conserved quantity, correlation functions of observables
overlapping with the conserved quantity will converge to a non-zero value. In the
RP resonance language, this would imply the presence of $\lambda_i = 1$ in the
decomposition in Eq.~\eqref{eq:rp_decay_show}. Additionally, also all powers of
the conserved quantity are conserved quantities, although typically non-local.
While they cannot cause correlation functions of extensive (or local)
observables to converge to a non-zero value in the thermodynamic limit, they
can introduce plateaus in finite systems that decay as a power law in system
size. We discuss this effect in Appendix~\ref{app:plateaus}.

\subsection{Example: Brickwall quantum circuits with no conserved quantities}
\label{sec:bw_ex}

\begin{figure}[t!]
    \centering
    \includegraphics[width=0.45\linewidth]{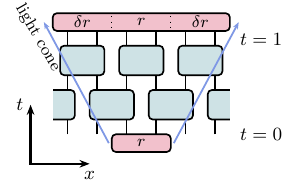}%
    \raisebox{110pt}{\hspace{-200pt}(a)\hspace{195pt}}\\
    \includegraphics[width=0.40\linewidth]{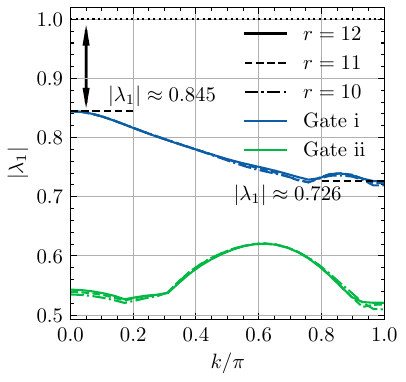}
    \raisebox{155pt}{\hspace{-175pt}(b)\hspace{170pt}}
    \includegraphics[width=0.415\linewidth]{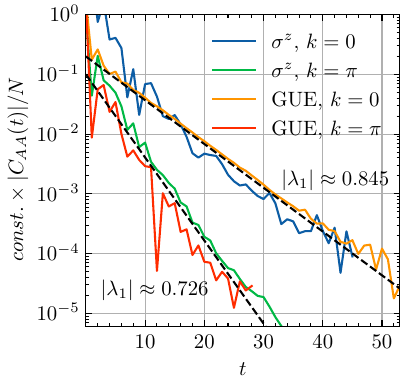}%
    \raisebox{155pt}{\hspace{-175pt}(c)\hspace{170pt}}
    \caption{Circuits with no conserved quantities. (a) Diagram of operator spreading in a
    brickwall circuit with the same gate $V$ acting between all nearest
    neighbors (the shown circuit has $s = 2$, $\delta r=2$). (b) The leading eigenvalue of the propagator
    truncated to the space of extensive observables with support $r$ for two
    \diff{realizations of a} circuit \diff{with a different gate $V$} chosen
    randomly according to the unitary Haar measure. (c) Infinite temperature
    autocorrelation functions of extensive
    observables $A$ \eqref{eq:a_ext_simple} in the circuit \diff{with gate i from} (b). The local density $a$ is either $\sigma^z$, or chosen randomly according to the Gaussian unitary ensemble
    (GUE)~\cite{mehtaRandomMatrices2004} with support $r = 2$. Dashed lines show the RP
    prediction $\propto \abs{\lambda_1}^t$, while full curves are an exact
    calculation in a circuit with \diff{$N = 32$} qubits; the $\sigma^z, k = 0$
    case is multiplied by $15$ for better presentation.}
    \label{fig:rp_no_sym}
\end{figure}

Let us demonstrate the above RP formalism with a simple example of a circuit
with no conserved quantities. The main idea is to compare the exactly calculated
correlation function in a \diffa{large} finite system with
the asymptotic decay obtained from the truncated propagator. In addition, we will show that
the quasi-momentum dependence of $\lambda_1(k)$ can be non-monotonous,
specifically, the slowest decay can happen at nonzero $k$.

We take a brickwall circuit with nearest-neighbor 2-qubit gates, all being the
same. The gate \diffa{that is repeated in the circuit} is picked randomly according to the unitary Haar measure. In
Fig.~\ref{fig:rp_no_sym}b, we show the leading RP resonance $\abs{\lambda_1(k)}$
for two realizations. For details on
numerical methods see Appendix~\ref{app:num_spec}. The circuit's Floquet
propagator can be written as
\begin{align}
    U &= U_\mathrm{even} U_\mathrm{odd}, \label{eq:bw_prop}\\
    U_\mathrm{odd} &= V_{1, 2} V_{3, 4} \dots V_{N - 1, N}, \nonumber \\
    U_\mathrm{even} &= V_{2, 3} V_{4, 5} \dots V_{N - 2, N - 1} V_{N, 1}, \nonumber
\end{align}
where $V_{i, j}$ are local gates, i.e., 2-site unitary operators acting on sites
$i$ and $j$. All $V$ are taken to be the same. In both \diff{realizations of a circuit} we see that the $k$ dependence
of $\lambda_1(k)$ is highly non-trivial and depends strongly on the choice of
$V$. \diff{A simillarly complicated behavior was also observed for the kicked Ising
model in Ref.~\cite{znidaricMomentumdependentQuantumRuellePollicott2024}.} We
show the $k$ dependence only for $k \in [0, \pi]$, since $\abs{\lambda_1(k)}
= \abs{\lambda_1(-k)}$. This is a consequence of the fact that $\left(\mathcal
U^{(r)}_k\right)^* = \mathcal U^{(r)}_{-k}$, which follows from the form in
Eq.~\eqref{eq:prop_mel}. One must only show that $\sprod{\mathcal S^{sj
+ m}(b)}{\mathcal U \left(\mathcal S^{m'}(b')\right)}$ is a real number, which
quickly follows from the hermiticity of the Pauli basis, $b = b^\dagger, b' =
b'^\dagger$.

Fig.~\ref{fig:rp_no_sym}c shows the decay of
correlation functions of extensive observables with $k = 0$ and $\pi$. For details about numerical calculation of correlation functions see Appendix~\ref{app:num_cf}. Their long-time
decay rate matches well with the leading RP resonance determined
from the truncated propagator. If the observables do not have good
quasi-momentum $k$, for instance strictly local observables, their correlation
function expands over the whole spectrum of eigenvalues with different momenta
$k$. The long time decay is then
governed by the biggest $\lambda_1(k)$. It is often assumed that this happens at
$k = 0$, i.e., for translationally invariant observables, but as we see in
Fig.~\ref{fig:rp_no_sym}b for \diff{the circuit with gate} ii,
this is not always the case. There the gap seems to be the smallest around
\diff{$k \approx 0.6 \pi$}.

In the following sections we shall focus on the main subject of our work,
namely, study how a conserved quantity changes the RP spectrum and what can we
extract from it. As mentioned in Sec.~\ref{sec:rp_res}, RP resonances can also
be studied by introducing Lindbladian
dissipation~\cite{moriLiouvilliangapAnalysisOpen2024}, though we find the
truncated propagator more convenient; we briefly discuss the Lindblad approach
in Appendix~\ref{app:lindblad}.

\section{The leading Ruelle-Pollicott resonance}
\label{sec:leading}

We now focus on the main result of this paper, the effect of a single conserved
quantity on the leading RP resonance. In order to study this, we need a model
with only 1 conserved quantity. A natural choice for the conserved quantity in the context of
spin-$\tfrac{1}{2}$ chains is the magnetization
\begin{equation}
    M \coloneq \sum_j \sigma_j^z.
\end{equation}
Therefore, our system, in addition to a translational symmetry, also has a U(1)
symmetry. A natural way to achieve that would be to take a brickwall circuit
with a 2-qubit gate that conserves magnetization. However, if the gate is the
same everywhere, such circuits are all special -- they are
integrable~\cite{znidaricIntegrabilityGenericHomogeneous2024,kpzwall}, including
for non-brickwall geometries~\cite{palettaIntegrabilityChargeTransport2025}.
The second possibility that also preserves translational symmetry would be to take a
brickwall configuration with two different magnetization-conserving gates, one
in each brickwall layer, though it turns out that such circuits have slow
(possibly non-diffusive)
thermalization~\cite{jonaySlowThermalizationSubdiffusion2024}.

\begin{figure}[t!]
    \centering
    \includegraphics[width=0.8\linewidth]{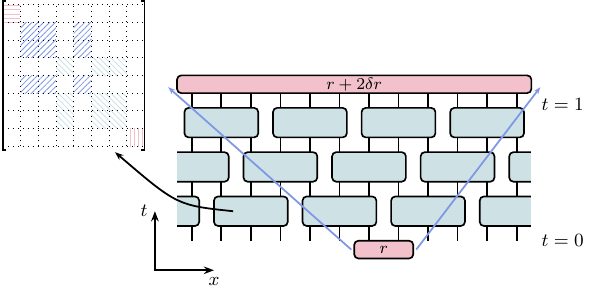}%
    \raisebox{152pt}{\hspace{-365pt}(a)\hspace{360pt}}
    \includegraphics[width=0.75\linewidth]{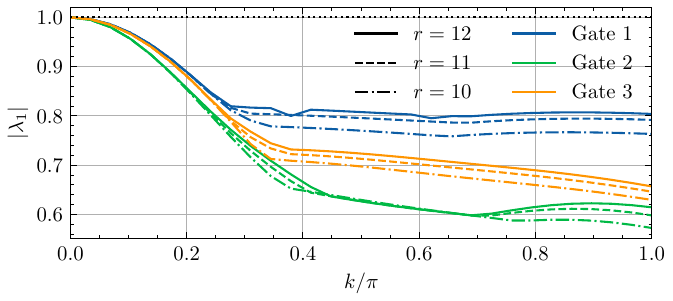}%
    \raisebox{140pt}{\hspace{-320pt}(b)\hspace{315pt}}
    \caption{Circuits with conserved magnetization. (a) Circuit diagram with
    magnetization\diffa{-conserving} 3-qubit gate $V$ (the shown circuit has $s = 3$ and
    $\delta r=6$), whose block structure is shown on the left.
    (b) The leading RP eigenvalue of the truncated propagator with support $r$
    for 3 choices of $V$ (chosen according to the Haar measure). The same \diff{gates}
    will be used throughout the paper.}
    \label{fig:rp_sym}
\end{figure}

Because we want good diffusive thermalization, we therefore consider one of the
next simplest options, which are qubit circuits with 3-site gates. The gates are
arranged in a \diffa{3-layer} brickwall-like geometry with $s = 3$ site translational
invariance, see Fig.~\ref{fig:rp_sym}a. The Floquet propagator is given by the
following expression
\begin{align}
    U &= U_3 U_2 U_1, \label{eq:3_prop}\\
    U_i &= \prod_{j = 0}^{N/3} V_{i + 3j, i + 3j + 1, i + 3j + 2} \nonumber
\end{align}
with periodic boundary conditions (indices are taken modulo $N$). 3-site gates
$V$ are taken the same everywhere, so that we have translational
symmetry.  In order for the circuit to conserve magnetization, \diffa{the} local gate \diffa{$V$} must
conserve it as well. One can thus quickly deduce that the local gate $V$ must
have 4 independent blocks, which are shown on the diagram in
Fig.~\ref{fig:rp_sym}a. \diffa{To get a typical representative of such
circuits}, we choose the matrix elements in each block independently
according to the unitary Haar measure in the corresponding block, which results in a unitary $V$.

The considered circuit spreads observables by $\delta r = 6$ in one period
(see Fig.~\ref{fig:rp_sym}) and, importantly, we have $\delta r > s$. Therefore,
Eq.~\eqref{eq:prop_mel} for the matrix elements of the truncated propagator is
not applicable. While one can use the more general formula derived in
Appendix~\ref{app:prop}, this approach gives access only up to $r = 6$, since
the evaluation of matrix elements is \diffa{performed} numerically by \diffa{the evolution
of} observables \diffa{on}
$r + 2\delta r$ \diffa{sites}. A substantial improvement can be made by noticing that
successive layers are translations of the previous layer, that is $\mathcal
S(U_i) = U_{i + 1}$. This space-time
symmetry~\cite{duhClassificationSamegateQuantum2024} allows us to redefine the
first layer (and a translation to the left) to be the ``new'' Floquet propagator
for a three times shorter period, leading to less spreading $\delta r = 2$, allowing
us to use Eq.~\eqref{eq:prop_mel} \diffa{and thereby enabling us to reach
about $r = 13$}. For details see
Appendix~\ref{app:space_time}.

The leading eigenvalue of the truncated propagator (and thus the leading RP
resonance) for $3$ different realizations with different gate $V$ is
shown in Fig.~\ref{fig:rp_sym}b. We will use the same \diff{gates} throughout this
paper. Its $k$ dependence has two regimes, $k$ close to $0$ (small $k$) and $k$
far from zero (around $k = \pi$). At $k = 0$, the leading eigenvalue is exactly
$1$ and its eigenvector is the local density corresponding to magnetization
(i.e., the conserved quantity \diffa{$a = \sigma^z$}). Eigenvalues at small quasi-momenta have smooth
dependence on $k$ and converge fast with $r$. We will show that they are related
to transport and contain information about the spin diffusion constant in
Sec.~\ref{sec:small_k}. Eigenvalues far from $k = 0$ converge slower and show
more varied behavior, similarly to what we have seen in \diff{circuits} without symmetries
(cf.~Fig.~\ref{fig:rp_no_sym}b), including non-smooth dependence due to
collisions with smaller eigenvalues. In Sec.~\ref{sec:big_k} we will numerically
verify that a finite gap there indeed signals exponential decay of corresponding
observables that are unrelated to transport.

\subsection{Small quasi-momenta and magnetization transport}
\label{sec:small_k}

In generic, i.e., chaotic, quantum systems with 1 conserved quantity, one
typically expects diffusive
transport~\cite{bertiniFinitetemperatureTransportOnedimensional2021}. Diffusive
transport can be probed by \diffa{looking at} the correlation function of the local
density of the conserved quantity, also called the dynamic structure factor.
\diffa{At sufficiently large $t$, it is expected to obey the diffusion
equation~\cite{forsterHydrodynamicFluctuations1990}, which gives}
\begin{equation}
    \ev{\sigma^z(x, t) \sigma^z(0, 0)} = \frac{1}{\sqrt{4 \pi D t}} \e{-\frac{x^2}{4Dt}}, \label{eq:diffusion}
\end{equation}
where $D$ is the model-dependent (spin) diffusion constant. We have adjusted our
notation to $\sigma^z(x, t)$ meaning the $z$-Pauli matrix acting on space
coordinate $x$ and propagated to time $t$. In Eq.~\eqref{eq:diffusion}, we consider
space to be continuous, with unit distance $\Delta x = 1$ corresponding to three sites (i.e., the
translational period) of the model.


Assuming Eq.~\eqref{eq:diffusion} holds, one can derive the correlation function of magnetization with quasi-momentum $k$, $M_k
\coloneq \sum_j \e{-\ii k j/3} \sigma_j^z$ [factor $3$ is included to match
the definition in Eq.~\eqref{eq:a_ext}]. That is,
\begin{align}
    \frac{1}{N}&\ev{M_k(t) M_k(0)^\dagger} = \frac{1}{N}\sum_{j, j'} \e{-\ii k (j - j')/3}\ev{\sigma^z_j(t) \sigma^z_{j'}(0)} \nonumber \\
        &= \sum_j \e{-\ii k j/3}\ev{\sigma^z_j(t) \sigma^z_0(0)} \nonumber\\
        &\rightarrow \int \ev{\sigma^z(x, t) \sigma^z(0, 0)} \e{-\ii k x} \dd{x} = \e{-D k^2 t}. \label{eq:mag_k_corr}
\end{align}
The correlation function of magnetization in momentum space therefore decays
exponentially, with the decay time diverging as $1/(Dk^2)$ in the long
wavelength limit $k \to 0$. This is a standard indicator of
diffusion~\cite{forsterHydrodynamicFluctuations1990}. Because there are
evidently exponentially decaying correlation functions with their rate having
quadratic $k$ dependence, this implies that there must exist an RP resonance
\begin{equation}
    \abs{\lambda_1(k)} = \e{-D k^2}. \label{eq:diff_rp_res}
\end{equation}
This can alternatively be interpreted as an RP continuum governing
the decay of the local magnetization correlation function. We discuss other
potential RP continuums in greater detail in Sec.~\ref{sec:hydro_modes}.

\begin{figure}[t!]
    \centering
    \begin{tikzpicture}
        \node[anchor=south west,inner sep=0] (image) at (0,0) {\includegraphics[width=0.45\linewidth]{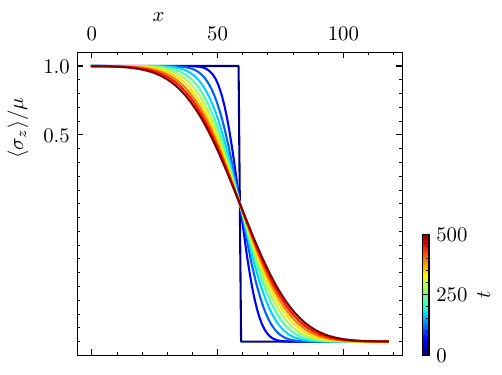}};
        \begin{scope}[x={(image.south east)},y={(image.north west)}]
            \node[anchor=south west,inner sep=0] (image) at (-0.085,-0.25) {\includegraphics[width=0.25\linewidth]{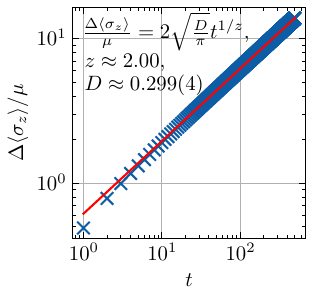}};
            \node[anchor=south west,inner sep=0] (image) at (0.5,0.4) {\includegraphics[width=0.255\linewidth]{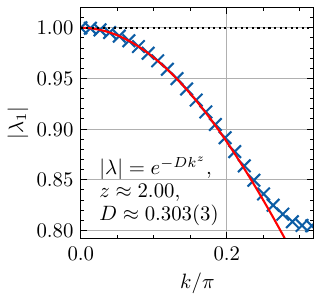}};
            \node[anchor=south west] at (0, 0.85) {(a)};
            \node[anchor=south west] at (0.9, 0.91) {(a.i)};
            \node[anchor=south west] at (0.23, 0.02) {(a.ii)};
        \end{scope}
    \end{tikzpicture}
    \hspace{-19pt}
    \begin{tikzpicture}
        \node[anchor=south west,inner sep=0] (image) at (0,0) {\includegraphics[width=0.45\linewidth]{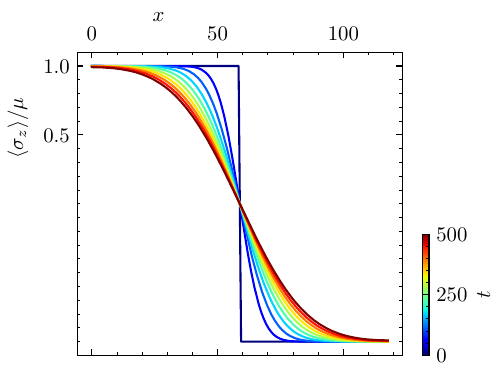}};
        \begin{scope}[x={(image.south east)},y={(image.north west)}]
            \node[anchor=south west,inner sep=0] (image) at (-0.085,-0.25) {\includegraphics[width=0.25\linewidth]{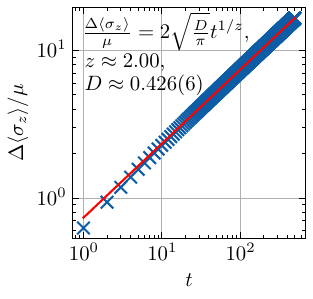}};
            \node[anchor=south west,inner sep=0] (image) at (0.5,0.4) {\includegraphics[width=0.245\linewidth]{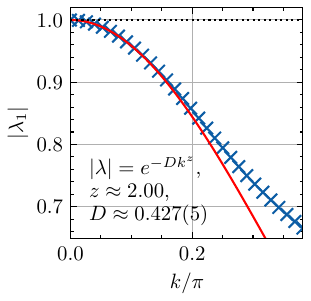}};
            \node[anchor=south west] at (0, 0.85) {(b)};
            \node[anchor=south west] at (0.85, 0.9) {(b.i)};
            \node[anchor=south west] at (0.23, 0.02) {(b.ii)};
        \end{scope}
    \end{tikzpicture}
    \caption{Transport of magnetization in the circuit with \diff{gate} 1 (a) and \diff{gate} 2 (b).
    In both \diff{instances} we show the leading RP resonance ($r=11$) and the fit to
    $\e{-Dk^z}$ (insets i), and the domain-wall evolution (TEBD, \diff{$N=354$} with
    $\mu = 10^{-3}$ and $\chi = 256$) at times with equal spacing \diff{$\Delta t =
    55$ up to $t = 500$} (main plots) with the associated transferred
    magnetization and the diffusion constant fit (insets ii).}
    \label{fig:transport}
\end{figure}
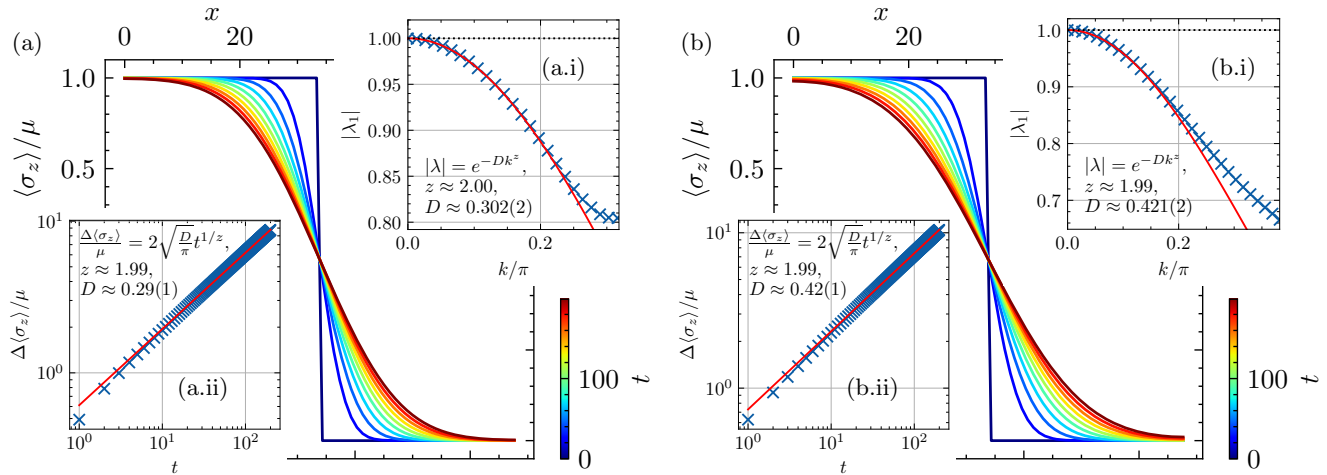

Since the RP resonance predicted in Eq.~\eqref{eq:diff_rp_res} converges to $1$
as $k \to 0$, it must be the leading RP resonance for small $k$. It is,
therefore, precisely the RP resonance observed for small $k$ in
Fig.~\ref{fig:rp_sym}b. This is confirmed numerically by fitting a function
$\e{-D k^z}$ to $\lambda_1$ for small $k$, where $z$ is usually referred to as
the transport dynamical exponent. This is shown in Fig.~\ref{fig:transport},
insets (i). The obtained values of $z$ match the predicted diffusive $z = 2$ up
to accessible precision. One can also check the corresponding eigenvectors,
which, according to the calculation in Eq.~\eqref{eq:mag_k_corr}, should overlap
with the local density of magnetization. We have confirmed that this is
\diffa{indeed} the case. It would be interesting to extend this \diffa{to a type
of non-diffusive transport} ($z \neq 2$) with a one-parameter scaling relation
$x^z \sim t$. We expect the quasi-momentum dependence of $\lambda_1(k)$ for
small $k$ to also hold information about $z$ in that case, although the exact
form of $\lambda_1(k)$ depends not just on $z$, but also on the scaling function
of the local correlation function of the conserved quantity.

The fit also extracts the spin diffusion constant $D$ of the model. In order to
verify that this is indeed the correct value, we compare it to the diffusion
constant obtained from a completely different method, namely, from the evolution
of a weakly-polarized domain-wall
state~\cite{ljubotinaSpinDiffusionInhomogeneous2017}. We prepare the system in
the state with the following density matrix
\begin{equation}
    \rho \propto \prod_{j = 1}^{N/2} \e{\mu \sigma_j^z} \prod_{j = N/2 + 1}^N \e{-\mu \sigma_j^z}, \label{eq:rho}
\end{equation}
for small $\mu$, and evolve it with $U$. The diffusion equation dictates that
the transferred magnetization $\Delta \ev{\sigma^z}_\rho$ should be (after long enough
time)
\begin{equation}
    \Delta \ev{\sigma^z}_\rho(t) \coloneq \sum_{j=N/2 + 1}^N \ev{\sigma_j^z(t) \diff{- \sigma_j^z(0)}}_\rho = 2 \mu \sqrt{\frac{D}{\pi} t}, \label{eq:delta_s}
\end{equation}
where $\ev{\bullet}_\rho \coloneq \tr(\rho\,\bullet)$ denotes the expectation value in
\diffa{the} state $\rho$. The dynamics can be simulated using time-evolving block decimation
(TEBD)~\cite{vidalEfficientClassicalSimulation2003,vidalEfficientSimulationOnedimensional2004}, for details see
Appendix~\ref{app:num_tebd}. Fitting a curve scaling as $t^{1/z}$ to $\Delta
\ev{\sigma^z}$ one can numerically confirm diffusion and extract the spin
diffusion constant $D$.

Both methods of determining $D$ are shown in Fig.~\ref{fig:transport}. In the
transferred magnetization from the domain wall (bottom left (ii) insets), we
again see $z$ matching the expected $z = 2$. The diffusion constants determined
from different approaches also match up to accessible precision \diffa{of
about $1 \%$}. We have thus
shown the leading RP resonance for small $k$ truly does contain information
about transport. A natural next question, which we leave for future work, is how
competitive is the extraction of $D$ through RP resonances compared to the
current state-of-the-art.

\subsection{Quasi-momenta far from zero and correlation function decay}
\label{sec:big_k}

\begin{figure}[t!]
    \centering
    \includegraphics[width=0.225\linewidth]{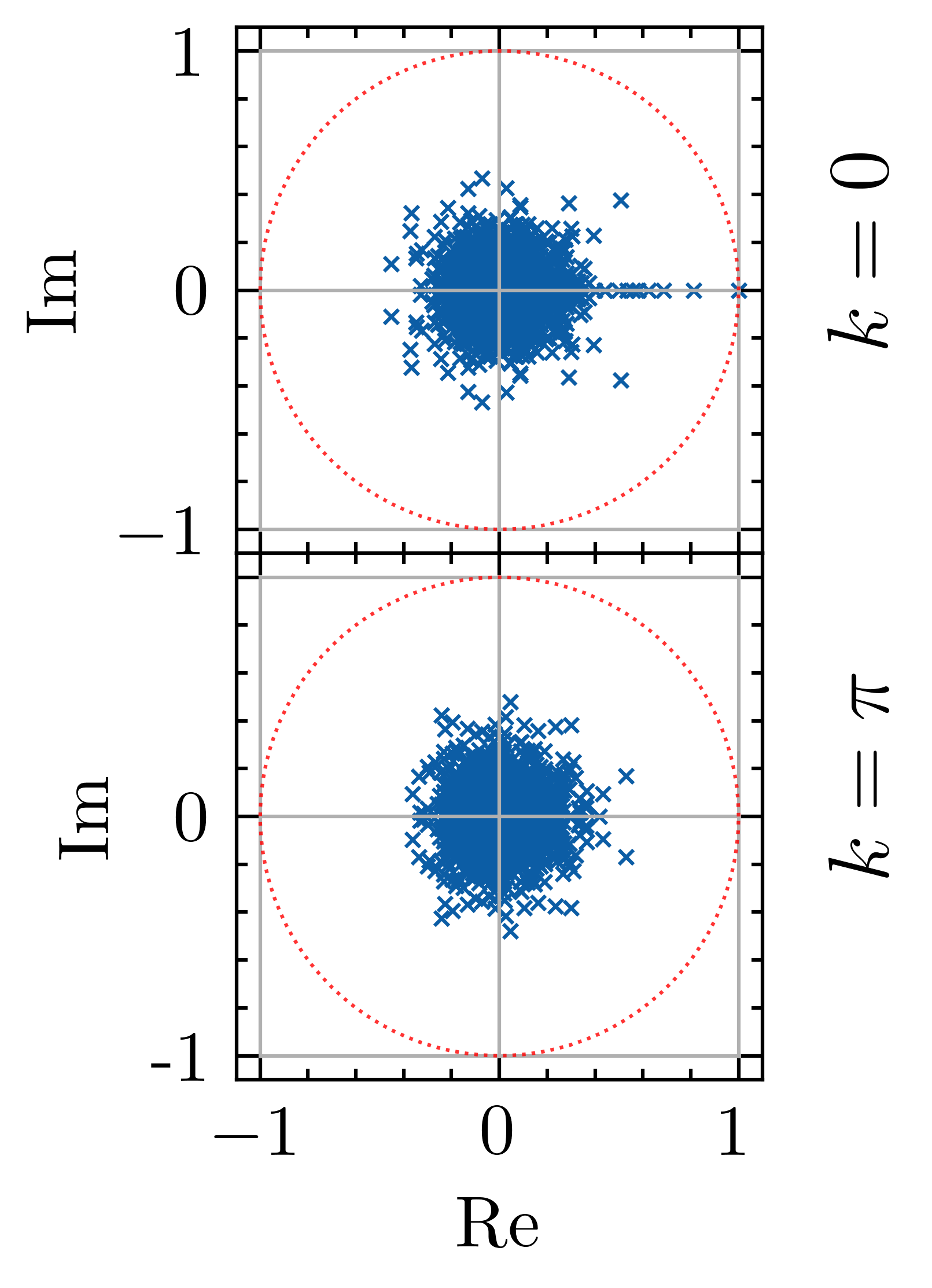}
    \raisebox{130pt}{\hspace{-102pt}(a)\hspace{84pt}}
    \includegraphics[width=0.373\linewidth]{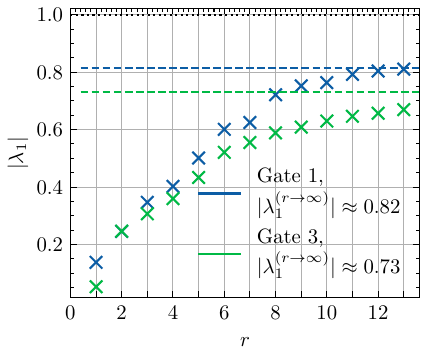}
    \raisebox{130pt}{\hspace{-167pt}(b)\hspace{152pt}}
    \includegraphics[width=0.3855\linewidth]{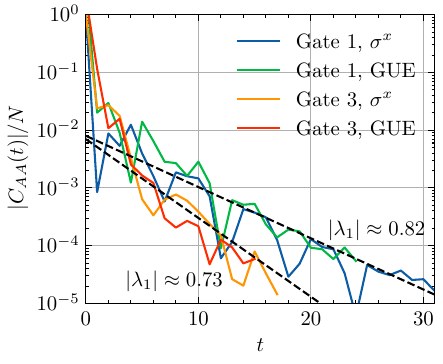}
    \raisebox{130pt}{\hspace{-167pt}(c)\hspace{147pt}}
    \caption{Leading RP resonance and correlation function decay in the $k =
    \pi$ sector for magnetization-conserving circuits. (a) The whole spectrum
    of the truncated propagator in $k = 0, \pi$ sectors for \diff{the circuit with gate} 3 and $r = 7$.
    (b) Convergence of the leading RP resonances in the $k = \pi$ sector with
    $r$, with \diffa{dashed lines denoting} an extrapolation to the $r \to \infty$ limit. (c) Infinite
    temperature autocorrelation functions of extensive observables $A$ defined
    in Eq.~\eqref{eq:a_ext} with $k = \pi$. The local densities are either
    $\sigma^x$ (i.e., $x$ magnetization) or chosen randomly according to the
    Gaussian unitary ensemble (GUE)~\cite{mehtaRandomMatrices2004} with support $r = 2$. The correlation
    functions are calculated in a finite circuit with $N = 30$ sites. Dashed
    lines show the RP predictions $\propto \abs{\lambda_1}^t$.}
    \label{fig:rp_sym_decay}
\end{figure}

The leading RP resonances far from $k = 0$ do not follow $\e{-Dk^2}$ anymore
and are not connected to diffusive transport, see Fig.~\ref{fig:rp_sym}b. We thus
expect them to merely be indicators of exponential decay, like in the case without
symmetries in Fig.~\ref{fig:rp_no_sym}. We demonstrate this numerically for $k
=\pi$ in Fig.~\ref{fig:rp_sym_decay}. Convergence with $r$ is slower than at
small $k$ and at the largest available $r=13$ the leading eigenvalue has not yet fully converged,
however, it does seem that the extrapolated values\footnote{The extrapolation
is done through an exponential fit. We have no good theoretical justification
for the exponential fit, it was simply the best match for the data.} are
strictly smaller than $1$. Therefore, generic correlation functions are expected
to decay exponentially with a finite rate. In Fig.~\ref{fig:rp_sym_decay}c we
show correlation functions of $\sigma^x$ magnetization and of a random \diffa{2-site} observable.
While exponential decay at available system size $N=30$ is not super nice, the
decay is compatible with the extrapolated $\lambda_1$, e.g., it is
faster for the circuit with smaller $\abs{\lambda_1}$. Also, different
observables decay with the same rate.

The fact that the exponential decay in the \diffa{shown circuits} is not as evident from
numerics is compatible with the slow convergence of the RP resonance with $r$.
Namely, the leading RP resonance at truncation $r$ can be intuitively understood
to govern the decay of correlations when the operator has spread to $r$ sites.
Slower convergence of RP resonances with $r$ thus implies ``nice'' exponential
decay only at later times and consequently smaller values of $C_{AA}$.
Therefore, it is expected that exponential decay will be harder to see
numerically at finite $N$ precisely when RP resonances will also be hard to
extract numerically. Additionally, correlation functions in
Fig.~\ref{fig:rp_sym_decay}c visibly oscillate, making numerical checks
less clear. This, however, is merely a consequence of $\lambda_1$ having an
imaginary component, see, e.g., \diff{gate} 3 in
Fig.~\ref{fig:rp_sym_decay}a.

\section{Beyond the leading Ruelle-Pollicott resonance}
\label{sec:subleading}

We have shown that the leading RP resonance $\lambda_1(k)$ for small $k$ is
connected to transport of the conserved quantity. How about subleading
eigenvalues; they could contain information about other correlation functions.
One might expect that correlation functions of observables orthogonal to all
transport-related quantities decay exponentially, like in a system without any
conserved quantities, which would imply that there must exist subleading RP
resonances.

\begin{figure}[t!]
    \centering
    \includegraphics[width=0.57\linewidth]{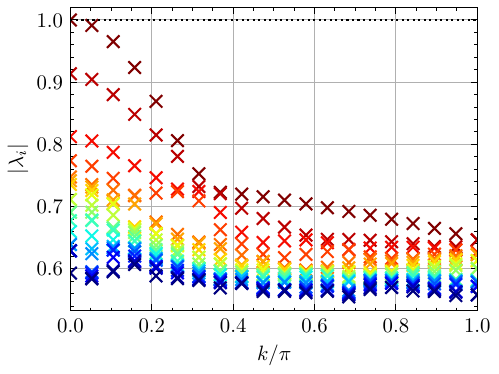}
    \caption{20 leading eigenvalues of the truncated propagator at different
    $k$. Calculated for \diff{the circuit with gate} 3 and support \diff{$r = 11$}. \diff{The
    subleading eigenvalues are not yet converged at accessible $r$, see the discussion
    in the main text and Fig.~\ref{fig:subleading_eigs} for details.}}
    \label{fig:subleading_eigs_k}
\end{figure}

The first few largest eigenvalues of the truncated propagator can be calculated
numerically and are shown in Fig.~\ref{fig:subleading_eigs_k}. We observe many
subleading eigenvalues in a narrow interval and one might think that they are
isolated RP resonances. It turns out this is not \diffa{the case}; these
eigenvalues have not yet converged with $r$ and, moreover, many of them look to
be converging to the leading eigenvalue $\lambda_1$ \diff{(see
Fig.~\ref{fig:subleading_eigs} for the convergence to $\abs{\lambda_1} = 1$ in
the $k = 0$ sector)}. We shall propose in Sec.~\ref{sec:powers_in_spec} that
in the $k= 0$ sector, some of such eigenvalues close to $1$ come from the powers of the
conserved magnetization $M$. Furthermore, we shall argue in
Sec.~\ref{sec:hydro_modes} that power-law hydrodynamic tails in
transport-related correlation functions imply an RP continuum in the spectrum of
the truncated propagator as sketched in Fig.~\ref{fig:rp_diagram}b. Both of
these results make it hard to numerically determine whether subleading discrete
RP resonances exist.

\subsection{Powers of the conserved quantity}
\label{sec:powers_in_spec}

Eigenvectors corresponding to $\lambda = 1$ are conserved quantities and while
our system does have only 1 local conserved quantity -- the magnetization
$M$ -- all its powers are also conserved. Because our
truncated operator propagator is geared towards local operators, while
$M^2=\sum_{i,k} \sigma_i^z \sigma_k^z$ is non-local, it does not have an
eigenvalue $\lambda=1$ that would correspond to $M^2$. However, as we will now
argue\footnote{We thank Rustem Sharipov for this suggestion.}, it does have an
eigenvalue that is close to $1$.

Without truncation, $M^2$ is conserved. This can be also seen explicitly,
${\mathcal U}(M^2)=$ $\sum_{i,k} {\mathcal U}(\sigma_i^z){\mathcal
U}(\sigma_k^z)$, which, after invoking definition of the spin current $j_k$ [see
Eq.~\eqref{eq:cont}], ${\mathcal U}(\sigma_k^z)=\sigma_k^z+j_{k-1}-j_{k}$, can
be rewritten as $\mathcal U(M^2) =
\sum_{i,k}(\sigma_i^z+j_{i-1}-j_i)(\sigma_k^z+j_{k-1}-j_k) =$ $M^2$. Conservation
comes about because the many-qubit terms $j_i j_k$ (and likewise $\sigma_i^z
j_k$) sum to $0$ due to alternating signs stemming from the continuity equation.

\begin{figure}[t!]
    \centering
    \includegraphics[width=0.6\linewidth]{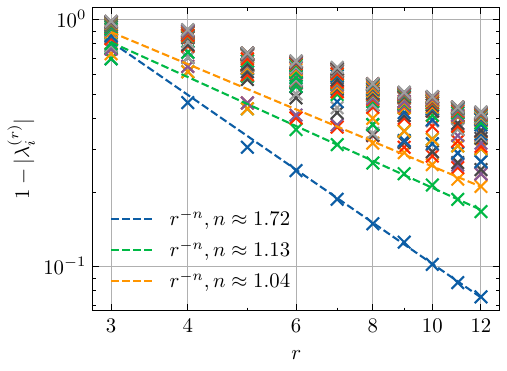}
    \caption{Convergence of 50 subleading eigenvalues of the truncated propagator in the $k =
    0$ sector to the unit circle. A power-law fit for the 3 subleading eigenvalues
    is shown.}
    \label{fig:subleading_eigs}
\end{figure}
\diff{In the basis truncated to support $r$, some of the terms needed for cancellation
are missing, resulting in an almost-conservation. Let us denote the
truncated $M^2$ by $Q^{(r)}$, explicitly $Q^{(r)}=\sum_{i,k,|k-i| < r}
\sigma_i^z \sigma_k^z$. In particular, the $j_i j_k$ terms coming from $\mathcal{U}(\sigma_i^z \sigma_k^z)$ with $\abs{k - i} = r - 1$, that would
cancel out with the same (but negative) terms from $\mathcal{U}(\sigma^z_{i} \sigma^z_{k +1})$, do not cancel out, since $\sigma^z_{i} \sigma^z_{k + 1}$ is not included in
$Q^{(r)}$ (and similarly for $\sigma_i^z j_k$ terms). Importantly, one can check that for $\abs{k -i} < r - 1$ the cancellation still occurs and all such terms are conserved. The upshot is that the only terms differing between $Q^{(r)}$ and $\mathcal U(Q^{(r)})$ are those coming from $\sigma_i^z \sigma_k^z$ with $\abs{k - i} = r - 1$, the number of which does not scale with $r$.

The number of terms in $Q^{(r)}$ is equal to the number of indices with $|k-i| <
r$, i.e., it scales as $\sim r$ ($\sim Nr$ in a system of length $N$). On the other hand, the number of term differing between $\mathcal U(Q^{(r)})$ and $Q^{(r)}$ is constant, $\sim c$. In the limit $r \to \infty$ the relative difference in small. Therefore, this suggests that $Q^{(r)}$ is almost an eigenvector with eigenvalue being close to $1$, which we estimate to be}\footnote{\diff{More exactly, given a
diagonalizable matrix $M$ and any vector $v$, we have $\norm{M v} = \norm{P^{-1}
\Lambda P v} \leq \abs{\lambda_v} \norm{P}_\mathrm{op} \norm{P^{-1}}_\mathrm{op}
\norm{v}$. Here, $\Lambda$ is the diagonal matrix of eigenvalues, $P$ the
transition matrix into the eigenbasis of $M$, $\norm{\bullet}_\mathrm{op}$ the
operator norm corresponding to $\norm{\bullet}$ and $\lambda_v$ the largest
eigenvalue (by magnitude) such that the corresponding eigenvector overlaps with
$v$. The inequality gives $\abs{\lambda_v} \geq \frac{1}{\norm{P}_\mathrm{op}
\norm{P^{-1}}_\mathrm{op}} \frac{\norm{M v}}{\norm{v}}$, which differs from
Eq.~\eqref{eq:lam_pred} only by the factor of $\frac{1}{\kappa(M)} =
\frac{1}{\norm{P}_\mathrm{op} \norm{P^{-1}}_\mathrm{op}}$. $\kappa(M)$ is the
spectral condition number and measures the non-normality of a matrix (for normal
matrices $\kappa = 1$, for non-normal $\kappa > 1$). For the truncated
propagator one can numerically check that $\kappa$ is large and increases with
$r$, making the bound uninformative. Nevertheless, the numerical evidence in
Fig.~\ref{fig:subleading_eigs} suggests that the informal
estimate (\ref{eq:lam_pred}) still roughly holds. This can be traced back to the empirical fact that the subleading eigenvector is well behaved and almost equal to $Q^{(r)}$.}}
\begin{equation}
    \lvert \lambda \rvert \approx 1-\frac{\norm{\mathcal{U}\!\left(Q^{(r)}\right)-Q^{(r)}}}{\norm{Q^{(r)}}} = 1 - \frac{c}{r}. \label{eq:lam_pred}
\end{equation}
We expect this eigenvalue also in the spectrum of the truncated propagator $\mathcal U^{(r)}_{k = 0}$.
A similar informal calculation can be done for an arbitrary power of magnetization.

Numerical tests of Eq.~\eqref{eq:lam_pred} by calculating $50$ subleading
eigenvalues of the truncated propagator are shown in
Fig.~\ref{fig:subleading_eigs}. A few largest subleading eigenvalues indeed
converge to the unit circle according to a power law. The first subleading
eigenvalue $\lambda_2$ in fact seems to converge even faster than the predicted
$1/r$. This might be a consequence of small $r$ accessible numerically.
Further analysis reveals that the eigenvector corresponding to $\lambda_2$ has a
large overlap with $Q^{(r)}$. Similarly, the second subleading eigenvector
(corresponding to $\lambda_3$) seems to have substantial overlaps with the
truncation of $M^3$.

The truncated propagator in $k=0$ therefore seems to have a number of
eigenvalues that converge to $1$ as $1/r$ that come from powers of $M$. With the
truncation that we use it is, therefore, hard to numerically see if there are
any genuine isolated subleading resonances. Furthermore, it is not clear if
subleading RP resonances should even exist due to power-law hydrodynamic tails
in correlation functions, as we discuss in the next section.

\subsection{Hydrodynamic tails}
\label{sec:hydro_modes}

In systems with conserved quantities, certain correlation functions related to
transport exhibit power-law decays, i.e., the so-called hydrodynamic
tails~\cite{forsterHydrodynamicFluctuations1990,mukerjeeStatisticalTheoryTransport2006},
due to slow power-law spreading of inhomogeneities of the conserved density
(distance and time scaling as $x^z \sim t$). A typical example is the
correlation function of local magnetization (\ref{eq:diffusion}), i.e., the
Green's function of the diffusion equation, decaying as $\sim
1/t^{1/2}$ at $x = 0$. Similar power-law tails occur also in higher-order corrections to
diffusion
equation~\cite{forsterHydrodynamicFluctuations1990,mukerjeeStatisticalTheoryTransport2006},
more recently studied in the context of effective field
theories~\cite{doyonDiffusionSuperdiffusionHydrodynamic2022,
matthies2024thermalizationhydrodynamiclongtimetails,
crossleyEffectiveFieldTheory2017,
michailidisCorrectionsDiffusionInteracting2024}.

\begin{figure}[t!]
    \centering
    \includegraphics[width=0.50\linewidth]{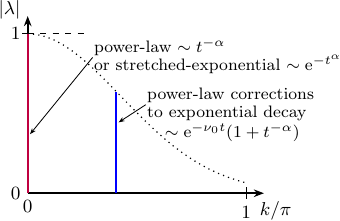}
    \caption{Diagrams of possible RP continuums and the type of
    correlation function decay of extensive observables they govern.}
    \label{fig:continuums_sketch}
\end{figure}

Power-law decay $C(t) \sim t^{-\alpha}$, for some $\alpha > 0$, being slower
than exponential, cannot be governed by a discrete RP resonance. If we
nonetheless assume it expands over a spectrum of eigenvalues as $C(t) = \sum_j
c_j \lambda_j^t$, where $c_j$ are expansions coefficients
[cf.~Eq.~\eqref{eq:rp_decay_show}], it must expand over a continuum of
eigenvalues (an ``RP continuum'') with closing gap to $\lambda =
1$\footnote{The argument can be illustrated more explicitly by writing the
decomposition as an integral $C(t) = \sum_j c_j \e{-\nu_j t} = \int_0^\infty
c(\nu) \e{-\nu t} \dd{\nu}$, where $\lambda = \e{-\nu}$, $c(\nu)$ is the
continuous version of the expansion coefficients and we assumed $\lambda$ to be real
for simplicity. In other words, $C(t)$ is the Laplace transform of $c(\nu)$. By
employing the inverse Laplace transform, we see that $C(t) = t^{-\alpha}$ is
\diffa{obtained for} $c(\nu) = \frac{1}{\Gamma(\alpha)} \nu^{\alpha - 1} \theta(\nu)$,
where $\Gamma$ is the Gamma function and $\theta$ is the Heaviside step
function. We observe that $c(\nu) \neq 0$ for all $\nu > 0$ -- an RP continuum
over all $\lambda \in [0, 1]$.}. This is precisely what we observed for the
correlation function of magnetization in momentum space \eqref{eq:mag_k_corr},
with the gap closing at $k=0$. Additionally, the quasi-momentum dependence of
the RP continuum was exactly determined in that case, $\abs{\lambda(k)} =
\e{-Dk^2}$. The expansion over eigenvalues for all $k$ is expected only for
correlation functions of local observables. For extensive observables, the whole
continuum must lie in one $k$ sector. A potential RP continuum governing a
power-law decay of a $k = 0$ extensive observable is depicted in the diagram in
Fig.~\ref{fig:continuums_sketch}.

Power-law decay is not the only type of non-exponential decay expected in
systems with one conserved quantity. One other possibility is a power-law
correction to exponential decay $C(t) \sim t^{-\alpha} \e{-\nu_0 t}$, for some
$\alpha$ and $\nu_0 > 0$, which is governed by a continuum $\lambda \in [0,
\e{-\nu_0}]$\footnote{$C(t) = t^{-\alpha}\e{-\nu_0 t}$ leads to $c(\nu) =
\frac{1}{\Gamma(\alpha)} (\nu - \nu_0)^{\alpha - 1} \theta(\nu - \nu_0)$ via the
inverse Laplace transform.}, also depicted in Fig.~\ref{fig:continuums_sketch}.
Such decays were observed in corrections to the correlation function of
magnetization with quasi-momentum $\ev{M_k(t) M_k(0)}$ (see
Eq.~\eqref{eq:mag_k_corr} for the exponential diffusive prediction), recently
explicitly calculated with the methods of effective field
theory~\cite{michailidisCorrectionsDiffusionInteracting2024}. The final
possibility is a stretched-exponential decay $C(t) \sim \e{-t^\alpha}$ for some
$0 < \alpha < 1$, which was recently predicted to occur for all
local correlation functions unrelated to
transport in 1d~\cite{mccullochSubexponentialDecayLocal2025}. It is governed
by a continuum $\lambda \in [0, 1]$ that we discuss further in
Appendix~\ref{app:stretched_exp}.

Based on the above arguments, we conjecture that there is an RP continuum of
eigenvalues for all $|\lambda|$ below the leading RP resonance governing the
transport of magnetization. In other words, all $|\lambda| \leq \e{-D k^2}$ are
part of an RP continuum.

\subsubsection{Extensive spin current}
\label{sec:current}

Power-law tails are expected also in the spin
current $j$ defined via the (discrete space and time) continuity equation
\begin{equation}
    \mathcal U(M_{[i, l]}) - M_{[i, l]} = j_{i - 1} - j_l, \label{eq:cont}
\end{equation}
where $j_i$ is the current operator between sites $i$ and $i + 1$, and we
introduced the magnetization acting on sites $i$ to $l$ as $M_{[i, l]} \coloneq
\sum_{j = i}^l \sigma^z_j$. The closed-form expression for $j_i$ is complicated
to derive analytically for our 3-layer brickwall circuit, but can be obtained
numerically for a given gate by evaluating the LHS of Eq.~\eqref{eq:cont} in a
big enough finite circuit. Current densities $j_i$ are different for different
values of $i$ modulo $3$, and have support on at most $9$ sites. For a
2-layer brickwall circuit (e.g., Fig.~\ref{fig:rp_no_sym}a) the expressions are
more manageable (2 or 4 site operators) and are written out for any
magnetization-conserving gate in Ref.~\cite{kpzwall}.

The asymptotics \mzz{at large $t$} of the local spin current correlation function can be deduced
from the local magnetization correlation function and the continuity equation.
The diffusive local magnetization correlation function~\eqref{eq:diffusion}
(i.e., the diffusion equation Green's function) implies
\begin{align}
    \label{eq:jj}
    \ev{j(x, t) j(0, 0)} = \frac{x^2 - 2 D t}{8\sqrt{\pi D} t^{5/2}} \e{-\frac{x^2}{4Dt}}.
\end{align}
For the derivation see Appendix~\ref{app:current}. At $x=0$ this again decays as
a power-law $\sim 1/t^{3/2}$.

Of particular interest is the extensive current
\begin{equation}
    J \coloneq \sum_i j_i,
\end{equation}
because it appears in the Green-Kubo equation for the diffusion
constant~\cite{kuboStatisticalPhysicsII1991}\footnote{For discrete-time dynamics, the Green-Kubo formula reads $D =
\lim_{t_\mathrm{max} \to \infty} \lim_{N\to \infty}\frac{1}{N} \big(\frac{1}{2}
\ev{J(0) J(0)} +$ $\sum_{t = 1}^{t_\mathrm{max}} \ev{J(t) J(0)}\big)$. In the
considered circuits, it gives $D \approx 0.31$ in \diff{the circuit with gate} 1 and $D \approx 0.44$ in
\diff{the circuit with gate} 2, which matches our predictions with other methods up to accessible
precision (see Fig.~\ref{fig:transport}).}.
Using Eq.~\eqref{eq:jj} and the fact that the equilibrium correlation function is
translationally invariant in space, we get \mzz{for the leading order of} the extensive current in the thermodynamic
limit
\begin{align}
    \lim_{N \to \infty}\frac{1}{N} \ev{J(t) J(0)} = \int_{-\infty}^\infty \ev{j(x, t) j(0, 0)} \dd{x} = 0. \label{eq:j_k_cf}
\end{align}
The \mzz{long-time} prediction coming from the diffusive (i.e., first order)
local magnetization correlation function, thus gives exactly $0$, which can be
traced back to the fact that $\ev{j(x, t)j(0, 0)}$ is a total derivative in $x$
[see Eq.~\eqref{eq:jj}]\footnote{\mzz{Note that, interestingly, according to the
Green-Kubo formula, the diffusion constant $D$ must acquire its nonzero value at
times before the local magnetization correlation function (\ref{eq:diffusion})
gets its Gaussian form and $\ev{J(t)J(0)}$ becomes zero.}}. This is very
interesting as it means that the leading \diff{asymptotic} behavior of the
correlation function of the extensive current is not $\sim t^{-3/2}$ (in
1d), but is rather given by subleading corrections to the local magnetization
correlation function. Correlation function of $J$ is, therefore, a sensitive
probe of subleading corrections to diffusion, that would perhaps be hard to see
in other correlation functions that are nonzero already in the leading order.

\begin{figure}[t!]
    \centering
    \includegraphics[width=0.60\linewidth]{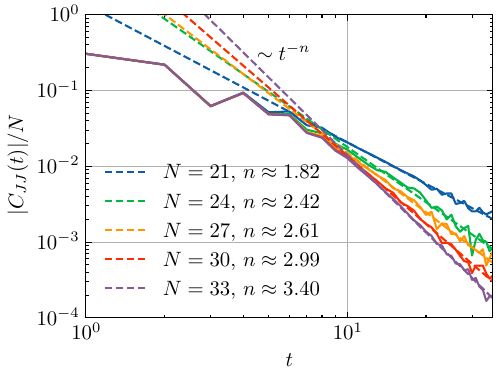}
        \caption{Infinite temperature autocorrelation of extensive spin current. The
        correlation functions are calculated in \diff{circuits with gate} 3 with
        a different number of sites $N$. Dashed lines show power-law fits.}
    \label{fig:current}
\end{figure}

Numerical calculation of the extensive current correlation function for one
\diff{circuit realization} is shown in Fig.~\ref{fig:current}. It decays as a
power law for any finite system size $N$, however, the power increases with $N$.
It is not clear what will happen in the thermodynamic limit; its power-law
dependence may converge to some finite power or the decay may become
exponential.

In addition to the inconclusive numerics, it is also not clear what is
theoretically expected. Power-law decay $\sim t^{-3/2}$ is expected in systems
with an additional conserved quantity, e.g., in Hamiltonian
systems~\cite{mukerjeeStatisticalTheoryTransport2006}, and can be explained by
the cross transport of the conserved quantities. In systems with only one U(1) conserved
quantity, a power-law decay with a higher power was
predicted~\cite{michailidisCorrectionsDiffusionInteracting2024}, although
exponential decay is also sometimes observed, e.g., in the kicked Ising
model~\cite{yi-thomasComparingNumericalMethods2024}.
\diff{The power law predictions are based on the corrections to the diffusive
local magnetization correlation function calculated from the effective field
theory of diffusion~\cite{crossleyEffectiveFieldTheory2017}, however,} \mzz{from
the calculation in Ref.~\cite{michailidisCorrectionsDiffusionInteracting2024}, it
is not clear what is the first nonzero correction contributing to the extensive
current correlation function. Furthermore, subleading corrections can depend on
unknown model-dependent parameters.}

All in all, the fundamental question of the asymptotics of the extensive current
correlation function $C_{JJ}$ still remains. Is it a power law, like the local
current correlation function, or are the hydrodynamic tails absent and, if they
are, when and why?

\section{Conclusion}

We have studied the (leading) Ruelle-Pollicott resonances in translationally invariant
qubit circuits \diffa{with 3-site magnetization-conserving gates}. \diffa{The
resonances were obtained from} the (leading)
eigenvalues of the truncated operator propagator in the limit of no truncation.
Our main result is that the quasi-momentum dependence of the leading
Ruelle-Pollicott resonance $\lambda_1$ can be used to extract the transport
properties. Specifically, for diffusive systems, we have $\abs{\lambda_1(k)} =
\e{-Dk^2}$ for small \diffa{quasi-momenta} $k$, allowing us to numerically determine the spin
diffusion constant $D$. For large $k$, the leading resonance is not related to transport
and governs the decay of respective correlation functions. We also argue for the
existence of a continuum of eigenvalues below the leading diffusive resonance,
$\abs{\lambda} < \e{-Dk^2}$, which controls non-exponentially decaying
correlation functions, for instance, due to power-law hydrodynamic tails.

Even though our analysis was limited to a specific qubit circuit model, we
expect our conclusions about the Ruelle-Pollicott resonances to be generic for
systems with exactly one U(1) conserved quantity. In particular, the final
expression for the matrix elements of the truncated propagator in
Eq.~\eqref{eq:prop_mel} [or, in the case of larger operator spreading,
Eq.~\eqref{eq:prop_mel_gen}] hold for any quantum circuit, for example, with
a different arrangement of gates or higher local dimension. An interesting open
question is generalizing the truncated propagator to Hamiltonian systems.

Another compelling direction is benchmarking the truncated propagator method for
extracting the diffusion constant against other standard methods. While we have
shown that the Ruelle-Pollicott method produces the same results as a
tensor-network simulation and the Green-Kubo formula, it is not clear whether
and when one method might be preferable to others\footnote{\mzz{While a detailed
analysis would be nice, it seems that the RP approach has some advantages
compared to the Lindblad-based approaches with dissipation targeting long Pauli
strings, like the DAOE~\cite{PhysRevB.105.075131}. Namely, we work directly in the
infinite-size limit and do the truncation instantly and explicitly. For
instance, for a particular system in Fig.~7 of Ref.~\cite{prethermal} the RP
approach gives the diffusion constant that is within less than $1\%$ of what is
believed to be the correct value, while the DAOE is about $3\%$ off (and the
Krylov method about $5\%$ off).}}. Additionally, the leading Ruelle-Pollicott
resonance can
hold information about non-diffusive transport, i.e., ballistic, sub- or
super-diffusive. However, the exact dependence of the leading resonance on $k$
depends on the asymptotic (space and time) behavior of the correlation function
of the local density of the conserved quantity (i.e., the scaling function) and
might be complicated. This can hinder the effectiveness of identifying the
transport type and extracting model-dependent constants. Especially interesting
are integrable systems with an extensive number of conserved quantities, where
we expect an extensive number of RP continuums governing transport.

A focus of future work can also be to better understand the conjecture about
subleading Ruelle-Pollicott continuums. A promising direction seems to be
developing a different truncation scheme that separates asymptotically
differently behaving observables, i.e., those which are governed by
hydrodynamics and those which are not. This can also shed additional light on
the stretched-exponential decay recently predicted for correlation functions of
local non-hydrodynamic observables in 1d
systems~\cite{mccullochSubexponentialDecayLocal2025}. A particularly interesting
physical question in this context is also the asymptotic behavior of the
extensive current correlation function. We have shown that  it is governed by the
corrections to the diffusive correlation function of local magnetization
recently argued to be universal for systems with exactly one U(1) conserved
quantity~\cite{michailidisCorrectionsDiffusionInteracting2024}.

\section*{Acknowledgements}
U.D. would like to thank Rustem Sharipov, Matija Koterle, Lenart Zadnik, and Tomaž
Prosen for insightful discussions. We also acknowledge support by Grants No.
J1-4385 and No. P1-0402 from the Slovenian Research Agency (ARIS).

\pagebreak
\appendix

\section{Details about the quasi-momentum-dependent propagator}

In this appendix, we more carefully define the mathematical structure of the
space of extensive observables briefly explained in Sec.~\ref{sec:prop} and use
it to derive the matrix elements of the truncated propagator,
Eq.~\eqref{eq:prop_mel}.

\subsection{Space of extensive observables}
\label{app:ext_obs}

As explained in words in Sec.~\ref{sec:prop}, the space of extensive observables
of support $r$ is spanned by
\begin{equation}
    \mathbb P_{k}^{(r)} \coloneq \bigcup_{m = 0}^{s - 1} \left\{B_k^{(m, b)} \ | \ b \in \mathcal P^{(r)}\right\}, \label{eq:ext_basis}
\end{equation}
where $B_k^{(m, b)}$ is defined in Eq.~\eqref{eq:ext_b} and $\mathcal P^{(r)}$ in
Eq.~\eqref{eq:local_b}. $\mathbb P_{k}^{(r)}$ is an orthonormal basis w.r.t.\ the extensive Hilbert-Schmidt
inner product, $\sprode{A}{B} = \frac{s}{N 2^N}\tr A^\dagger B$, defined in Sec.~\ref{sec:prop}. This can be
deduced from a direct calculation,
\begin{align}
    &\sprode{B^{(m, b)}_k}{B^{(m', b')}_k} = \\
    &= \frac{s}{N 2^N} \tr \left[\sum_{j} \e{\ii k j} \mathcal S^{sj + m}(b) \sum_{j'} \e{-\ii k j'} \mathcal S^{sj' + m'}(b') \right] = \nonumber \\
    &= \frac{s}{N 2^N}\sum_{j, j'} \e{\ii k (j - j')} \tr  \left[\mathcal S^{sj + m}(b) \mathcal S^{sj' + m'}(b')\right] = \nonumber \\
    &= \frac{s}{N} \sum_{j, j'} \e{\ii k (j - j')} \delta_{m, m'} \delta_{j, j'} \delta_{b, b'} = \delta_{m, m'}\delta_{b, b'}, \nonumber
\end{align}
where $\delta$ is the Kronecker delta. We used that
\begin{equation}
    \sprod{\mathcal S^j(b)}{\mathcal S^{j'}(b')} = \delta_{j, j'} \sprod{b}{b'} \label{eq:s_loc_zero}
\end{equation}
for $b, b' \in \text{Span}(\mathcal P^{(r)})$. Here the local
Hilbert-Schmidt inner product is used, $\sprod{a}{b} = \frac{1}{2^N}\tr
a^\dagger b$, defined already in Sec.~\ref{sec:prop}. This is a consequence of the fact that $b, b'$ act with a
traceless operator on the first site. For any $j \neq j'$ we thus get a
traceless operator acting on either site $j$ or $j'$, rendering the whole inner
product zero. If additionally $b, b' \in \mathcal P^{(r)}$, we have
$\sprod{b}{b'} = \delta_{b, b'}$.

\subsection{Propagator of extensive observables}
\label{app:prop}

We now express the Heisenberg propagator $\mathcal U$ in the basis $\mathbb
P_{k} \coloneq \lim_{r \to \infty} \mathbb P_{k}^{(r)}$. Since we are
working with an orthonormal basis, the matrix elements $\left[\mathcal
U_k\right]_{(m, b),(m', b')}$ are simply
\begin{align}
    &\left[\mathcal U_k\right]_{(m, b),(m', b')} = \sprode{B^{(m, b)}_k}{\mathcal U\left(B^{(m', b')}_k\right)} = \nonumber \\
    &= \frac{s}{N} \sum_{j, j'} \e{\ii k (j - j')} \sprod{\mathcal S^{sj + m}(b)}{\mathcal U \left(\mathcal S^{sj' + m'}(b') \right)} \nonumber \\
    &= \frac{s}{N} \sum_{j, j'} \e{\ii k (j - j')} \sprod{\mathcal S^{s(j - j') + m}(b)}{\mathcal U \left(\mathcal S^{m'}(b')\right)} \nonumber \\
    &= \sum_{j} \e{\ii k j} \sprod{\mathcal S^{sj + m}(b)}{\mathcal U \left(\mathcal S^{m'}(b')\right)}, \label{eq:prop_mel_gen}
\end{align}
where we took into account translational invariance and the cyclic property of
trace.

In many cases, the sum in Eq.~\eqref{eq:prop_mel_gen} has only a few non-zero terms.
The reason for that is similar to the one employed in showing
Eq.~\eqref{eq:s_loc_zero}: since $b, b'$ act with a traceless Pauli operator on the first site,
their translations must overlap on it in order to give a non-zero inner product.
In addition to that, Eq.~\eqref{eq:prop_mel_gen} contains propagation $\mathcal U\left(
\mathcal S^{m'} (b')\right)$, which can change the first site on which $b'$ acts
non-trivially.

As mentioned already at the beginning of Sec.~\ref{sec:trunc_prop}, in quantum
circuits, the support of observables can increase by at most a finite amount in
one period, i.e., the light cone is sharp. In general, the spreading of
operators depends on the geometry, they can even spread to all sites in one time
period, e.g., in staircase circuits. Additionally, it depends on the site at
which they act, e.g., a two site operator in a brickwall circuit shown in
Fig.~\ref{fig:rp_no_sym} spreads by $1$ site to the left and right if it begins
at an odd site and by $2$ sites to the left and right if it begins at an even
site (the case depicted in the figure). Furthermore, it can also be different to
the left and to the right, e.g., in the circuit considered in the bulk of this
paper, the operator shown in Fig.~\ref{fig:rp_sym} spreads by $6$ sites to the
left and $4$ sites to the right. To simplify the discussion, we define $\delta
r$ to be the maximum spreading for any site and any direction (left/right) in a
considered circuit. One can easily see that if $\delta r \leq s$, only three
terms in Eq.~\eqref{eq:prop_mel_gen} can be non-zero, and we can simplify it to
\begin{equation}
    \left[\mathcal U_k\right]_{(m, b), (m', b')} = \sum_{j = -1}^1 \e{\ii k j} \sprod{\mathcal S^{sj + m}(b)}{\mathcal U \left(\mathcal S^{m'}(b')\right)}. \label{eq:prop_mel_app}
\end{equation}
This is precisely Eq.~\eqref{eq:prop_mel} we set out to derive. An analogous
form was written down in particular cases in
Refs.~\cite{prosenChaosComplexityQuantum2007,
znidaricMomentumdependentQuantumRuellePollicott2024,
znidaricIntegrabilityGenericHomogeneous2024}. The sum has an interpretation of
``realigning'' propagated densities back into the basis, where the mapping
between densities and extensive observables is one-to-one. For details about
this interpretation see Appendix A in
Ref.~\cite{sharipovErgodicBehaviorsReversible2025}.

Eq.~\eqref{eq:prop_mel_app} applies to many circuits studied in the literature,
such as the brickwall circuits, where $\delta r = s = 2$, see
Fig.~\ref{fig:rp_no_sym}. It also applies to the whole set of canonical geometries
with 2-site nearest-neighbor gates introduced in
Ref.~\cite{duhClassificationSamegateQuantum2024}. In
the 3-site gate circuits considered in the bulk of this paper (see Fig.~\ref{fig:rp_sym}),
the spreading is faster, though; there $\delta r = 6$ and $s = 3$ and therefore
more than $3$ terms in Eq.~\eqref{eq:prop_mel_gen} can be non-zero. One can,
however, use a trick described in the following section.

\subsection{Space-time symmetries}
\label{app:space_time}

While matrix elements of the truncated propagator for arbitrary spreading
$\delta r$ can be evaluated by Eq.~\eqref{eq:prop_mel_gen}, the expressions can
sometimes be simplified by exploiting a space-time symmetry. Namely, in both the
brickwall circuit defined in Eq.~\eqref{eq:bw_prop} and the 3-site
generalization of the brickwall defined in Eq.~\eqref{eq:3_prop}, one can notice
that successive layers (i.e., $U_{\mathrm{odd/even}}$ or $U_{1/2/3}$) are
translations of the previous layer. This can be interpreted as a space-time
symmetry; translations in space are equivalent to translations by a fraction of
a period in time. In equation, valid for both of the mentioned cases,
\begin{equation}
    S^\dagger U(0, 1) S = U(1/s, 1 + 1/s),
\end{equation}
where $S$ is the 1-site translation operator to the left, $S^\dagger (a \otimes
\1) S = \1 \otimes a = \mathcal S(a \otimes \1)$, $s$ is the number of sites of translation invariance
(and, crucially, also the number of layers), and $U(t_1, t_2)$ denotes the
propagator from time $t_1$ to $t_2$. Time is defined to be propagated by $1/s$ after the
application of each layer, in particular, the Floquet propagator is $U \equiv
U(0, 1)$.

Space-time symmetries and their implications are discussed in greater detail in
Ref.~\cite{duhClassificationSamegateQuantum2024}. An important result for a large class
of circuits is that their
propagator can be written in the following way
\begin{equation}
    U = S^{-s} \left(S \tilde U\right)^s, \label{eq:space-time_prop}
\end{equation}
where $\tilde U$ is the Floquet propagator of a circuit of some other (preferably
simpler) geometry. In both of the considered examples, $\tilde U$ is the propagator of
the 1st layer, $\tilde U = U_\mathrm{odd}$ for the brickwall~\eqref{eq:bw_prop} and
$\tilde U = U_1$ for its 3-site generalization~\eqref{eq:3_prop}.

Since $S^{-s}$ is just a phase in the quasi-momentum eigenbasis, one can
redefine $S \tilde U$ to be the new equivalent Floquet propagator (i.e., the
propagator for ``one period''). In particular, for the 3-site generalization of the
brickwall, this reduces the spreading in one period from $\delta r = 6$ to
$\delta r = 2 < s = 3$, allowing us to use the simpler expression for the matrix
elements of the truncated propagator, Eq.~\eqref{eq:prop_mel}, and allowing for
bigger numerically reachable supports (it requires working with operators
supported on at most $r+2\delta r=r+4$ sites, instead of $r+12$), see
Appendix~\ref{app:num_spec} for details about the numerics. Since the long-time
behavior is not affected by the redefinition of the Floquet propagator, the
extracted RP resonances must be the same. One must merely account for the fact
that the new Floquet propagator $S \tilde U$ propagates only for time $1/s$ and
thus take the $s$-th power of the extracted resonances. Additionally, if one
would study the phase of RP resonances, they would have to be rotated in each $k$ sector
by the appropriate phase stemming from $S^{-s}$ in
Eq.~\eqref{eq:space-time_prop}.

\section{Numerical methods}
\label{app:num}


\subsection{Spectrum of the truncated propagator}
\label{app:num_spec}

RP resonances are extracted numerically from $\mathcal U_k^{(r)}$,
which can be represented as a finite matrix with matrix elements explicitly
given by Eq.~\eqref{eq:prop_mel}. Support $r$ basis has $\frac{3s}{4}4^r$
elements [see Eq.~\eqref{eq:ext_basis}], thus resulting in a $\frac{3s}{4}4^r
\times \frac{3s}{4}4^r$ matrix. Importantly, the evaluation of matrix
elements is performed by propagating observables with support $r + 2\delta
r$ and are, therefore, numerically represented as $\frac{3s}{4}4^{(r + 2\delta
r)}$ component vectors.

Exact diagonalization can be used for smaller supports, up to around $r = 7$ for
our case of $s = 3$. At larger supports, we are constrained by memory and are
forced to use iterative methods, which target only a finite number of
eigenvalues. Since we are typically interested only in the first few leading
eigenvalues, this can be done efficiently. We must only evaluate
Eq.~\eqref{eq:prop_mel} on the fly, i.e., by directly acting on a given
observable. The calculations are again done on observables with at most
support $r + 2 \delta r$, which means that the maximum reachable support is
constrained also by the amount of spreading. In the considered cases, this
approach gives us access to support around $r = 13$. We use Arnoldi iteration
implemented in ARPACK~\cite{arpack}.

The truncated propagator is generically a non-normal matrix (i.e., it does not
commute with its Hermitian adjoint) and its eigenvectors are, therefore, not
orthogonal. This can lead to numerical difficulties and subtleties when
interpreting its spectrum~\cite{trefethenSpectraPseudospectraBehavior2005}. It
was already observed in Ref.~\cite{prosenRuelleResonancesQuantum2002} that RP
resonance eigenvectors are singular objects, however, to what extent
non-normality causes other problems when extracting RP resonances remains to be
seen.


\subsection{Correlation functions}
\label{app:num_cf}

Throughout the paper we numerically calculate connected infinite temperature
autocorrelation functions, that is
\begin{align}
    C_{AA}(t) &= \ev{A(t)A(0)} = \frac{1}{2^N} \tr A(t)A(0) \label{eq:cf_trace}\\
    &= \frac{1}{2^N} \sum_{\psi} \mev{\psi}{A(t)A(0)}, \nonumber
\end{align}
where the sum is taken over some normalized basis of the Hilbert space, and $N$ is the
number of sites in the circuit. For clarity, we assumed traceless $A$, although
the following discussion straightforwardly generalizes to any $A$. Eq.~\eqref{eq:cf_trace}
can be numerically evaluated by definition up to $N = 18$. In bigger systems, we
are constrained by memory and must use the typicality approach. Namely,
fluctuations of $C_{AA}(t)$ in a single random state are of the order of $\sim
1/2^{N/2}$. Therefore, the trace in Eq.~\eqref{eq:cf_trace} can be approximated
by only 1 random state, resulting in error on the order of $10^{-5}$ at $N
= 33$. Crucially, both $A$ and $U$ must be sufficiently local for
this approach and their action on a state must be implemented efficiently. That
is, the full $2^N \times 2^N$ matrices must never be saved in memory, only their
action on a state must be evaluated on the fly [e.g., by Eqs \eqref{eq:a_ext},
\eqref{eq:bw_prop} and \eqref{eq:3_prop}].

\subsection{Domain wall quench}
\label{app:num_tebd}

To determine the diffusion constant independently of RP resonances, we simulate
unitary evolution starting with a weakly polarized domain wall state (see Sec.~\ref{sec:small_k}).
In order to probe the behavior in the thermodynamic limit, it is crucial that
the numerical simulation only goes up to times, where the boundary has not yet
meaningfully affected the dynamics. In diffusive systems, this time can be
approximated by the expected spreading of the diffusive front $t \sim
x_\mathrm{max}^2/D$. To reach big enough times, we use
TEBD~\cite{vidalEfficientClassicalSimulation2003,
vidalEfficientSimulationOnedimensional2004}. The diffusion constant is expressed
from the transferred magnetization in state defined in Eq.~\eqref{eq:rho}.
According to Eq.~\eqref{eq:delta_s}, this involves evaluating
\begin{equation}
    \ev{\sigma^z_j(t)}_\rho = \tr \rho(t) \sigma^z_j.
\end{equation}
There are multiple ways to numerically calculate this, for example one can write either
$\rho$ or $\sigma_z^j$ as a matrix product operator (MPO) and evolve them, or
appropriately sample and evolve pure states. For small $\mu$, evolving $\rho$
turns out to have the slowest growth of Schmidt coefficients and is therefore
the most efficient to simulate. This is compatible with previous findings in
similar systems, for details see
Ref.~\cite{ljubotinaSpinDiffusionInhomogeneous2017}. We used this method in the
main part of the paper setting the maximum bond dimension to $\chi
= 256$. Additionally, we checked the results
indeed converged by considering different $\chi$ and estimated the error
in the diffusion constant $D$ by fitting the prediction in Eq.~\eqref{eq:delta_s}
to different time windows.

\section{Ruelle-Pollicott resonances through weak dephasing}
\label{app:lindblad}

In this section, we demonstrate the weak Lindbladian dissipation-based method,
proposed in Ref.~\cite{moriLiouvilliangapAnalysisOpen2024}, on circuits without
any conservation, and briefly show it gives the same results as the truncated
propagator method. Additionally, we show that in the circuits we studied, the
convergence of the Lindblad method can sometimes be rather
tricky.

The idea is~\cite{moriLiouvilliangapAnalysisOpen2024} to introduce some
kind of Lindbladian dissipation of strength $\gamma$,
obtaining the open system version of the Heisenberg propagator $\mathcal
U_\gamma$ (which, in this case, is the exponent of the Lindbladian). The leading
RP resonances $\lambda$ then correspond to the leading eigenvalues of the
Heisenberg propagator $\lambda_{\mathcal U_\gamma}$ in the zero noise $\gamma
\to 0^+$ limit and in the thermodynamic limit, i.e.,
\begin{equation}
    \lambda = \lim_{\gamma \to 0^+} \lim_{N \to \infty} \lambda_{\mathcal U_\gamma},
\end{equation}
where $N$ is the system size. The order of limits is important, the opposite
order always gives $0$ gap.

What kind of dissipation to take is not entirely clear, likely some locality
is required. What we want to demonstrate is that the speed of convergence
of the above limits might be strongly dependent on the chosen form, or,
equivalently, on the chosen 2-qubit gate $V$ for a given dissipation. In this
paper we fix dissipation to dephasing and test circuits with different gates
$V$. More specifically, the considered system's Heisenberg propagator is
\begin{align}
    \mathcal D_\gamma(A) &\coloneq \gamma \sum_{j = 1}^N \left( \sigma^z_j A \sigma^z_j - A\right), \label{eq:open_circ}\\
    \mathcal U_\gamma(A) &\coloneq \e{\mathcal D_\gamma}(U^\dagger A U), \nonumber
\end{align}
where $U$ is the brickwall circuit Floquet propagator without magnetization
conservation defined in Eq.~\eqref{eq:bw_prop}. In other words, every time step
consists of first evolving with the brickwall circuit propagator and then
applying dephasing with strength $\gamma$ on every site.

\begin{figure}[t!]
    \centering
    \includegraphics[width=0.45\linewidth]{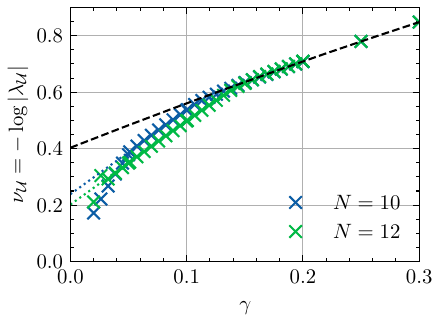}
    \raisebox{137pt}{\hspace{-195pt}(a)\hspace{185pt}}
    \includegraphics[width=0.438\linewidth]{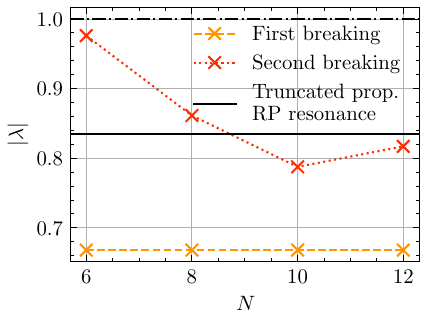}
    \raisebox{137pt}{\hspace{-190pt}(b)\hspace{170pt}}
    \includegraphics[width=0.47\linewidth]{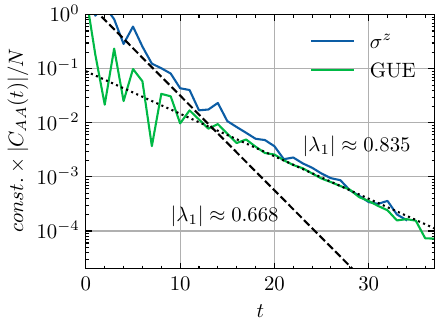}
    \raisebox{142pt}{\hspace{-203pt}(c)\hspace{183pt}}
    \caption{Extraction of the leading RP resonance by a weak-dephasing
    Lindbladian [see Eq.~\eqref{eq:open_circ}], an analogous figure to
    Fig.~\ref{fig:rp_no_sym}. (a) The leading Liouvillian eigenvalue for
    the dissipative brickwall quantum circuit with a Haar-random gate $V$. Dashed and dotted
    lines denote extrapolation to zero noise. (b) The
    convergence of the second breaking extrapolation [dotted lines in (a)] compared to
    the first breaking extrapolation [dashed lines (a)] and the RP resonance
    estimated from $r = 12$ truncated propagator (the biggest gap is at $k =
    0$). (c) Infinite temperature autocorrelation function of extensive $k = 0$
    observables, defined in Eq.~\eqref{eq:a_ext_simple}. The local densities are
    either $\sigma^z$ (i.e., $z$ magnetization) or chosen randomly according to
    the Gaussian unitary ensemble (GUE)~\cite{mehtaRandomMatrices2004} with support $r = 2$. The dashed and
    dotted lines denote the decays $\propto \abs{\lambda}^t$ governed by the
    first and second breaking predictions, respectively. The numerics are done
    in a finite circuit with \diff{$N = 32$} sites, the $\sigma^z$ correlation function
    is multiplied by $9$ for better presentation.}
    \label{fig:lindblad}
\end{figure}

One can calculate the leading eigenvalue of $\mathcal U_\gamma$ with an
iterative solver (similar to what we did for the truncated propagator,
described in Appendix~\ref{app:num_spec}) for different $N$ and $\gamma$.
Extrapolation to zero noise and the thermodynamic limit must be done
carefully~\cite{moriLiouvilliangapAnalysisOpen2024}. Namely, if one observes the
curve $|\lambda_{\mathcal U_\gamma}|$ for fixed $N$, we see that it ``breaks''
(i.e., abruptly changes its slope) at some small $\gamma$. This change has a
physical origin: for small $\gamma$ the scattering length becomes larger than
the system size and one starts to be influenced by finite-size (boundary)
effects. Therefore, in order to obtain the correct decay rate of the bulk
physics from finite-$N$ data one has to extrapolate the curve
$|\lambda_{\mathcal U_\gamma}|$ from finite (large) $\gamma$ down to $\gamma =
0$.

An example of such behavior is seen in numerical data for one realization of
a brickwall circuit (with a rather tricky instance of $V$) shown in Fig.~\ref{fig:lindblad}a. Instead of $\lambda$
we plot $\nu_{\mathcal U_\gamma} = - \log \abs{\lambda_{\mathcal U_\gamma}}$,
like in Ref.~\cite{moriLiouvilliangapAnalysisOpen2024}. We observe
$\nu_{\mathcal U_\gamma}$ is independent of the system size $N$ for big enough
$\gamma \gtrsim 0.15$. From data at $\gamma > 0.15$, we
can extrapolate (black dashed curve) to obtain the prediction for the leading RP
resonance $\abs{\lambda} \approx \e{-0.403} \approx 0.668$. Comparing this to the value obtained by
the truncated propagator (Fig.~\ref{fig:lindblad}b) and to decay of
correlation functions (Fig.~\ref{fig:lindblad}c), we see it is incorrect. The
long time decay is actually governed by $\abs{\lambda} \approx \e{-0.181} \approx 0.835$, which
the truncated propagator predicts correctly (in this case, this is the
maximum $\lambda_1(k)$ over all $k$, which occurs at $k = 0$).

If we study $\nu_{\mathcal U_\gamma}$ more carefully, we can get the
correct RP resonance from the Lindbladian approach too. Calculating
$\nu_{\mathcal U_\gamma}$ for smaller $\gamma$, we in fact observe another
breaking of the curve. This time, the curves for different $N$ do not overlap,
but we can still extrapolate for each $N$ separately and then try to do $N \to
\infty$. Such a procedure is also shown in Fig.~\ref{fig:lindblad}a (colored
dotted curves), with the convergence with $N$ shown in the inset (a.i). While
the procedure does not yet converge for accessible $N$, it is in the vicinity of
what appears to be the correct RP resonance as obtained from the truncated
propagator. Additionally, in Fig.~\ref{fig:lindblad} we see that the early time
behavior of the magnetization correlation function seems to be governed by
$\abs{\lambda} \approx 0.668$. The initial breaking of the $\nu_{\mathcal
U_\gamma}$ therefore might not be spurious, but rather an indication of a
subleading resonance.

We must emphasize that the showcased behavior is not generic. Double breaking is
not always observed; sometimes the correct RP resonance is given by the first
breaking, sometimes not even by the second. Additionally, we do not necessarily
see a 2-step relaxation in generic correlation functions. All in all, this
example showcases our experience that the Lindbladian method is less
robust. Taking the double limit involves making extrapolations that
are sometimes hard to perform. Furthermore, if we would want to treat systems
with a conservation law, like magnetization, the method is much more cumbersome,
and it is not clear how to do the momentum resolution efficiently.

\section{Plateaus in correlation functions due to the powers of the conserved quantity}
\label{app:plateaus}

Powers of a conserved quantity are also conserved quantities. When studying
physical properties of many-body systems, they can be often ignored because a
power of a local operator is a non-local operator. In finite systems of size
$N$, however, their impact can be non-negligible. Namely, local observables can
have a finite overlap with some power of a conserved quantity and thus cause its
correlation function to saturate to a higher value than the expected
fluctuations (see Appendix~\ref{app:num_cf}). Although this mechanism is simple,
we explicitly demonstrate it in the case of magnetization conservation, since
one needs to be aware of it when interpreting finite-size numerics.

Let $Q$ be a conserved quantity, i.e., $\mathcal U(Q) = Q$, which means its correlation
functions is constant,
\begin{equation}
    C_{QQ}(t) = \ev{Q(t) Q(0)} = \sprod{Q}{Q},
\end{equation}
where the angled brackets denote the Hilbert-Schmidt inner product,
$\sprod{A}{B} \coloneq \frac{1}{2^N}\tr A^\dagger B$. We additionally assumed
$Q$ to be Hermitian and traceless for simplicity, although a similar conclusion
can be made also in the general case.

If an observable overlaps with $Q$, $c_Q \coloneq \frac{\sprod{Q}{A}}{\sprod{Q}{Q}} \neq 0$, we also expect its
correlation function to be constant after a long enough time. Namely, we can decompose
$A$ in an orthogonal basis including $Q$, obtaining
\begin{align}
    C_{AA}(t) &= \ev{A(t)A(0)} = \sprod{c_Q Q(t) + \cdots}{c_Q Q(0) + \cdots} \nonumber \\
    &\xrightarrow[t \to \infty]{} \abs{c_{Q}}^2 \sprod{Q}{Q} = \frac{\sprod{Q}{A}^2}{\sprod{Q}{Q}}.
\end{align}
Here dots denote the decaying part, i.e, we assumed the system to be mixing and
that $Q$ is the only conserved quantity $A$ overlaps with. We also assumed $A$
to be traceless for simplicity.

Let's now demonstrate this with the correlation function of a $2$-site extensive observable
\begin{equation}
    A = \sum_j \sigma^z_j \sigma^z_{j + 1}
\end{equation}
evolved by a magnetization-conserving circuit. $A$ has zero
overlap with the conserved $M = \sum_j \sigma^z_j$, so its correlation
function should decay to $0$ in the thermodynamic limit $N \to \infty$. While
this is true, it has a non-zero overlap with powers of $M$ and its correlation
function will therefore exhibit a finite-size plateau.

Although $A$ overlaps with all even powers of magnetization, the leading order
of the plateau in $1/N$ will be caused by $M^2$. For $Q$ one
takes the traceless part of $M^2$, $Q = M^2
- N \1$. A short calculation gives
\begin{align}
    \sprod{M^2 - N \1}{A} &= 2 N, \\
    \sprod{M^2 - N \1}{M^2 - N \1} &= 2 N (N - 1).
\end{align}
Which results in the finite-size plateau
\begin{equation}
    \frac{1}{N} C_{AA}(t \to \infty) = \frac{1}{N}\frac{\sprod{Q}{A}^2}{\sprod{Q}{Q}} = \frac{2}{N - 1}. \label{eq:plateau}
\end{equation}
We show the result for the correlation function normalized by $1/N$, since
for extensive observables $C_{AA}(0) \propto N$.

\begin{figure}[t!]
    \centering
    \includegraphics[width=0.4748\linewidth]{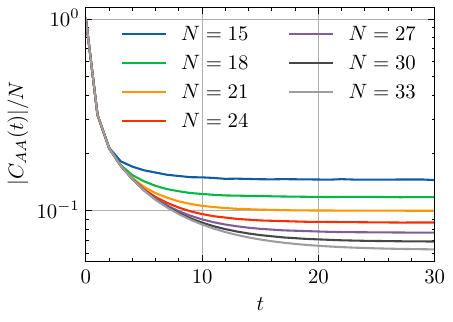}
    \raisebox{140pt}{\hspace{-206pt}(a)\hspace{193pt}}
    \includegraphics[width=0.50\linewidth]{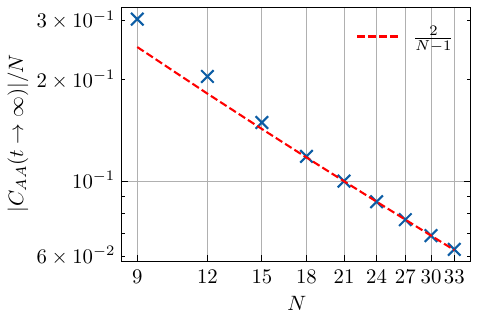}
    \raisebox{140pt}{\hspace{-218pt}(b)\hspace{198pt}}
    \caption{Finite-size plateaus in the autocorrelation function of $A = \sum_j
    \sigma^z_j \sigma^z_{j + 1}$. Shown are numerical results for a circuit with
    a random 3-site magnetization-conserving gate (Sec.~\ref{sec:leading}) and
    different $N$. Subfigure (a) shows time dependence, while (b) shows the
    values of plateaus in (a). The red dashed line are the leading asymptotics
    \eqref{eq:plateau}.}
    \label{fig:plateaus}
\end{figure}

The correlation functions of $A$ in a homogeneous circuit with a random 3-site gate (i.e.,
the one described in Sec.~\ref{sec:leading}) and different $N$ are shown in
Fig.~\ref{fig:plateaus}a. The plateaus are shown in Fig.~\ref{fig:plateaus}b and
match the predicted $\sim 1/N$ scaling at large $N$.

\section{Stretched-exponential decay of local correlation functions}
\label{app:stretched_exp}

Recently~\cite{mccullochSubexponentialDecayLocal2025}, it was argued that all
local correlation functions unrelated to transport decay as stretched
exponentials in diffusive one-dimensional systems, that is as $\sim \e{-B
t^\alpha}$ for some $B$ and $\alpha$. For translationally-invariant Floquet
systems, Ref.~\cite{mccullochSubexponentialDecayLocal2025} predicts $\alpha =
2/3$ and numerically confirms the prediction for magnetization-conserving
brickwall circuits.

\begin{figure}[t!]
    \centering
    \includegraphics[width=0.55\linewidth]{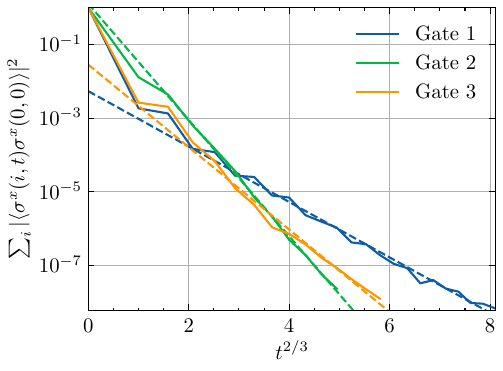}
    \caption{Local correlation function of $\sigma^x$ magnetization. The correlation
    function is calculated in a circuit with a random 3-site
    magnetization-conserving gate (Sec.~\ref{sec:leading}) with \diff{$N = 33$} sites
    and is averaged over space. Dashed lines show a stretched exponential $\e{-B
    t^{2/3}}$ fit.}
    \label{fig:stretched_exp}
\end{figure}

We check this in the 3-site generalized brickwall circuit (see
Sec.~\ref{sec:leading}) by analyzing the local correlation function of $\sigma^x$
magnetization, which is not expected to be described by an effective
(hydrodynamic) theory of transport~\cite{mccullochSubexponentialDecayLocal2025,
matthies2024thermalizationhydrodynamiclongtimetails}. The numerics are shown in
Fig.~\ref{fig:stretched_exp}, we average over space to obtain better statistics.
The average is done on the absolute value squared, since averaging just
$\ev{\sigma^x(i, t) \sigma^x(0, 0)}$ would yield the extensive $\sigma^x$ magnetization
correlation function. While the numerics shown suggest the decay is better
described by a stretched exponential with $\alpha = 2/3$ than an exponential,
fitting the power $\alpha$ has a big uncertainty (we obtain $\alpha \sim 0.4 -
0.8$ for different models with fitting uncertainties of the order of $\sim
0.1$).

As argued in Sec.~\ref{sec:hydro_modes}, stretched-exponential decays are
governed by RP continuums with closing gap. An interesting open question is what
happens for correlation functions of extensive observables, which are
essentially the Fourier transform of local observables. Our numerics seem to
indicate exponential decay, at least for big $k$ (see Sec.~\ref{sec:big_k}).
Asymptotics for correlation functions for $k = 0$ are less clear numerically,
the extensive $x$ magnetization does not seem to be in the asymptotic regime for
accessible time scales and additionally heavily oscillates.

\section{Local spin current}
\label{app:current}

In this section, we provide additional details about the diffusive behavior of
the local spin current discussed in Sec.~\ref{sec:current}. The spin current
$j$ is defined by the discrete continuity equation, Eq.~\eqref{eq:cont}. The diffusive
predictions come from the diffusive local correlation function of magnetization,
Eq.~\eqref{eq:diffusion}, and the continuity equation. For clarity, we use the
continuous space-time continuity equation
\begin{equation}
    \partial_t \sigma^z(x, t) = -\partial_x j(x, t),
\end{equation}
where $j(x, t)$ is the continuous version of the spin current from Eq.~\eqref{eq:cont}.
We derive
\begin{align}
    &\ev{j(x, t) j(0, 0)} = -\int_{-\infty}^x \dd{x'} \partial_t \ev{\sigma^z(x', t) j(0, 0)} \\
    &\qquad= -\int_{-\infty}^x \dd{x'} \partial_t \ev{\sigma^z(0, 0) j(-x', -t)} \nonumber \\
    &\qquad= \int_{-\infty}^x \dd{x'}\int_{-\infty}^{x'} \dd{x''} \partial_t^2 \ev{\sigma^z(0, 0) \sigma^z(-x'', -t)} \nonumber \\
    &\qquad= \int_{-\infty}^x \dd{x'}\int_{-\infty}^{x'} \dd{x''} \partial_t^2 \ev{\sigma^z(x'', t) \sigma^z(0, 0)}, \nonumber
\end{align}
where we used translational invariance and the cyclic property of trace. Plugging in the
diffusive magnetization correlation function~\eqref{eq:diffusion}, we obtain
\begin{align}
    \ev{j(x, t) j(0, 0)} = \frac{x^2 - 2 D t}{8\sqrt{\pi D} t^{5/2}} \e{-\frac{x^2}{4Dt}}.
\end{align}

Similarly, as in Sec.~\ref{sec:small_k}, the correlation function of extensive
current with good quasi-momentum $J_k := \sum_l \e{-\ii l k/3} j_l$ can now be
studied. We obtain
\begin{align}
    \frac{1}{N}\ev{J_k(t) J_k(0)^\dagger} &\rightarrow \int \ev{j(x, t) j(0, 0)} \e{-\ii k x} \dd{x} \\
    &\propto k^2 \e{-Dk^2 t}. \nonumber
\end{align}
The decay is, therefore, governed by an RP resonance with the same $k$ dependence
as the one governing the decay of magnetization [cf.~Eq.~\eqref{eq:mag_k_corr}],
with the prefactor $k^2$ perhaps interpreted as an expansion coefficient.

\pagebreak
\bibliography{references}

\end{document}